\documentclass[a4paper,11pt]{article}
\pdfoutput=1 

\usepackage{jheppub1} 

\usepackage[latin3]{inputenc}
\usepackage[makeroom]{cancel}
\usepackage{graphicx}
\usepackage{amsmath}
\usepackage{amsfonts}
\usepackage{amssymb}
\usepackage{color}
\usepackage[export]{adjustbox}
\usepackage{graphicx}
\usepackage{dcolumn}
\usepackage{bm}
\usepackage{simplewick}
\usepackage{array}
\usepackage{appendix}
\usepackage{relsize}
\usepackage{subcaption}
\usepackage{float}
\notoc

\newcommand{\be}{\begin{equation}}
	\newcommand{\ee}{\end{equation}}
\newcommand{\beq}{\begin{equation}}
	\newcommand{\eeq}{\end{equation}}
\newcommand{\bea}{\begin{eqnarray}}
	\newcommand{\eea}{\end{eqnarray}}

\title{\boldmath The non-Kerr black hole with acceleration}

\author[a, b]{Usman A. Gillani,}
\author[c]{Rehana Rahim}
\author[a]{and Khalid Saifullah}

\affiliation [a]{Department of Mathematics, Quaid-i-Azam University, Islamabad, Pakistan}
\affiliation [b]{National University of Technology, Islamabad, Pakistan}
\affiliation [c]{Department of Mathematics and Statistics, Riphah International University, Islamabad, Pakistan}

\emailAdd{usmangillani.qau@gmail.com}
\emailAdd{rehana.rahim8@gmail.com}
\emailAdd{ksaifullah@fas.harvard.edu}

\abstract{The no-hair theorem can be tested in the strong gravity regime by using the top-bottom approach and the bottom-top approach. The non-Kerr spacetime of the later approach is an ideal framework  to do the tests in the region very close to the black holes. In this work, we propose a non-Kerr black hole metric (and its charged extension) that is accelerating as well. These new objects are studied for their basic properties and thermodynamics. \vspace{80 mm}}

\begin{document}
\maketitle
\flushbottom
\section{Introduction}

The astrophysical black holes are described uniquely by  mass, charge and spin and are represented by the Kerr-Newman metric or the Kerr
metric in the absence of charge. This is the no-hair theorem.
Therefore, all the astrophysical black holes are supposed to be
described by the Kerr metric. One can test this black hole hypothesis in strong gravitational field by employing
 the top-bottom approach and the bottom-top approach \cite{bambi}. In the first case, there are
theoretical models in which the Kerr metric does not describe the spacetime of an astrophysical black hole \cite{car,kle}. The second approach
involves phenomenological modification of the metric. The metric in this approach can be some black hole solution in some theory of gravitation. There is no particular theory involved. The fundamental theory is unknown here, although it is assumed that particles will follow geodesics in this theory. The idea is to develop the understanding into the theory through observations. Many such spacetimes have been developed \cite {jp,tj,ch,ks,rla} from which probable observational signatures of the Kerr-like spacetimes can be studied. The degree of deviation from the Kerr varies from small to large in such Kerr-like spacetimes. The free parameters are employed to study the observable. Such modified spacetimes reduce to the Kerr metric when the new parameters are set to zero.

The Newman-Janis algorithm is an important solution generating technique  \cite{a}. Starting from a static spactime, this algorithm induces rotation in it. The Kerr and Kerr-Newman metrics have been derived by using this algorithm. The seeds metrics for the derivation of Kerr and Kerr-Newman black holes were taken as Schwarzschild and Reissner-Nordstr\"om metrics, respectively \cite {a,b}.

The non-Kerr metric proposed by Johannsen and Psaltis \cite{jp} is an important and extensively studied example of the bottom-top approach. 
It contains a set of free parameters besides  mass and spin. These parameters measure deviation from the Kerr metric. 
This metric is asymptotically flat, axisymmetric and vacuum solution of some unknown field equations which are different from the Einstein field equations for the non-zero deviation parameters. It has been shown that for the case of a single deformation parameter, this non-Kerr spacetime is regular and does not contain unphysical properties outside the event horizon and can represent a black hole upto the maximum range of the spin parameter. It is also an ideal framework for the tests of the no-hair theorem in the region very close to the black hole and which do not have explicit dependence on the field equations. The metric has been extended to include the electric charge and two different deviation functions \cite{jpc, cpr}.

The charged non-Kerr (deformed) black hole spacetime is given by \cite{jpc}
\begin{eqnarray}
ds^{2} &=&-(1+h(r,\theta ))\Big(1-\frac{2Mr}{\rho ^{2} }+\frac{q^2}{\rho ^{2} }\Big)dt^{2}  \notag \\
&&
-\frac{2a(2Mr-q^2)\sin
^{2}\theta }{\rho ^{2} }(1+h(r,\theta ))dtd\phi
+\frac{\rho ^{2} (1+h(r,\theta ))}{
\Delta +a^{2}\sin ^{2}\theta h(r,\theta )}dr^{2} +\rho ^{2} d\theta ^{2} 
\notag \\
&&
+\Big[\sin ^{2}\theta \Big(r^{2}+a^{2}+\frac{a^2(2Mr-q^2)\sin
^{2}\theta }{\rho ^{2} }\Big) 
+h(r,\theta )\frac{a^{2}\sin ^{4}\theta (\rho ^{2} +2Mr-q^2)}{
\rho ^{2} }\Big]d\phi ^{2}, \notag \\
&& \label{2.3a}
\end{eqnarray}
where
\begin{equation}
h=\frac{\epsilon M^3 r }{ \rho^4}.\label{h} \end{equation}
 Here $\epsilon$ is the deformation parameter, $a$, $M$, $q$ denote the spin, mass and charge of the black hole, respectively. It reduces to the non-Kerr spacetime in the case, $q=0$, and to the Kerr-Newman spacetime when the deformation $\epsilon$ is made to vanish. It is also valid in the super spinning cases in strong gravitational field.  This metric was developed with a desire to create an analogue of the Kerr-Newman metric of GR in the alternate theories of gravity.

Karl Shawarzchild and Johannes Droste gave a non-trivial solution of Einstein's field equations in 1916 for first time in history but this solution was misunderstood due to the presence of singularity. This confusion was cleared after the discovery of David Finklestein, who provided a comprehensive description of the  singularity as a surface which can only be traversed in one direction. These solutions were mathematically good but were not good enough physically. Levi-Civita extended this work and improved the solutions \cite{LC}. After him, Ehlers and Kundt \cite{NT, EW} classified the research work of Levi-Civita in A, B and C classes of metrics. The name of C-metric still exists in the literature and it was interpreted as the solution of accelerated black holes in 1970 by Kinnersely and Walker \cite{KW}. Later on, Pleba\'nski and Demia\'nski discovered a new spacetime in $1976$ which covers a large family of electro-vacuum type-D metrics in GR. Starting from a general form of the metric, Einstein-Maxwell field equations are employed to generate the solutions \cite{pd}. The accelerating and rotating black hole solution is an important member of the Pleba\'nski and Demia\'nski family of spacetimes. Here, acceleration of the black hole is measured by the parameter $\alpha$. The metric represents the gravitational field of a pair of uniformly accelerating Kerr-type black holes.

If the cosmological constant is taken to be zero, the charged accelerating black hole solution is \cite{cam,cam1,uk,mk, vac}
\begin{align}
ds^{2} &=\frac{1}{\Omega ^{2}}\Big\{{-}\big(\frac{Q}{\rho ^{2}}-\frac{a^{2}P\sin
^{2}\theta }{\rho ^{2}}\big)dt^{2}{+}\frac{\rho ^{2}}{Q}dr^{2}+\frac{\rho ^{2}}{P}
d\theta^{2}
 \nonumber \\
&{+}
\sin^{2}\theta\big(\frac{P(r^{2}+a^{2})^{2} }{\rho ^{2}}{-}\frac{Qa^{2}\sin
^{2}\theta }{\rho ^{2}}\big)\Big\} d\phi ^{2}{-}\frac{2a\sin ^{2}\theta
\left[P\left(a^{2}+r^{2}\right){-}Q\right]}{\rho ^{2}\Omega ^{2}}dtd\phi, \label{2.1a}
\end{align}
where
\begin{align}
\Omega &=1-\alpha r\cos \theta , \label{ab} \\
\rho ^{2}&=r^{2}+a^{2}\cos ^{2}\theta , \\
P &=1-2\alpha M\cos \theta +\alpha ^{2}\left(a^{2}+q^{2}\right)\cos ^{2}\theta , \\
Q &=(a^{2}+q^{2}-2Mr+r^{2})(1-\alpha ^{2}r^{2}).  \label{2.2a}
\end{align}
Here  $\alpha$ denotes the acceleration parameter. For the case where the cosmological constant $\Lambda\geq 0$, with $\alpha=0$, the spacetime gives a single
black hole, but for $\alpha\neq 0$ it gives a pair of causally separated black holes that
are accelerating away from each other in opposite directions \cite{cam1, uk, pd1}. For  $\Lambda <0$, the spacetime is a single black hole if $\alpha$ is small and a pair of black holes if it has a large value \cite{dl1,dl2,kb}. 
Putting  $\alpha=0$ gives the Kerr-Newman metric. Setting
 $\alpha=0=q$ yields the Kerr metric.

In this article, we develop non-Kerr black holes (charged and neutral), that are accelerating, in some alternate theory of gravity. Here, the main concern lies with the form of the metric and its subsequent properties. When the deformation vanishes, these spacetimes reduce to the accelerating black holes.

The black hole evaporation and thermal radiation theory \cite{Hak, Hak1} has a significant role in the development of theoretical physics. It has provided a connection among the fields of GR, quantum physics and thermodynamics. The laws of black hole thermodynamics are similar to the laws of ordinary thermodynamics i.e. these massive objects obey the standard laws of thermodynamics. The field of black hole thermodynamics is an extensively studied field and can be considered as a gateway to understanding the quantum nature of black holes. This article introduces two  new metrics and we have discussed their thermodynamic properties.

The article is arranged as follows: In Section \ref{metric}  non-Kerr metrics with acceleration are  given. The basic structure of these spacetimes is also studied in this section. In Section \ref{thermo2} thermodynamic analysis of the new metrics with acceleration is done. In section \ref{thermojp} the  thermodynamic properties of the non-Kerr black hole have been studied. The work is concluded in Section \ref{dis} . Throughout this work, we take $G=c=1$.

\section{Non-Kerr black hole spacetime with acceleration }\label{metric}
In this section we propose two non-Kerr accelerating black hole metrics and their certain basic characteristics. First is the charged non-Kerr accelerating black hole metric and the second one is the special case of the new metric in the case of zero electric charge. 
\subsection{The charged case}
The charged non-Kerr accelerating spacetime is proposed as 
\begin{align}
ds^{2} &=\frac{1}{\Omega ^{2}}\Big\{-\Big(\frac{Q}{\rho ^{2}}-\frac{a^{2}P\sin
^{2}\theta }{\rho ^{2}}\Big)\left(1+h\right)dt^{2}+\frac{\rho ^{2}\left(1+h\right)}{Q+a^2h\sin^{2} \theta}dr^{2}+\frac{\rho ^{2}}{P}%
d\theta ^{2}  \nonumber \\
&{+}\sin ^{2}\theta\Big(\frac{P(r^{2}+a^{2})^{2} }{\rho ^{2}}{-}\frac{Qa^{2}\sin
^{2}\theta \left(1+h\right)}{\rho ^{2}}\Big)\Big\} d\phi ^{2}{-}\frac{2a\sin ^{2}\theta
\left[P\left(r^{2}+a^{2}\right){-}Q\right]\left(1{+}h\right) }{\rho ^{2}\Omega ^{2}}dtd\phi,\nonumber \label{2.1} \\
\end{align}
where $\Omega$, $\rho$, $P$, $Q$ and $h$ are given in Eqs. (\ref{ab})-(\ref{2.2a}) and Eq. (\ref{h}), respectively. Putting  $\alpha=0$ gives metric (\ref{2.3a}). Setting $\alpha=0=h$ gives the Kerr-Newman black hole. This metric does not obey the usual Einstein-Maxwell equations
due to the presence of $h(r,\theta)$. So, we make the assumption that the above spacetime might be an electro vacuum solution to some unknown field equations which are different from the Einstein-Maxwell equations for non vanishing $h(r,\theta)$.
To confirm the validity this black hole metric, we study its basic structure, that is, event horizon, ergosphere and the redshift. 

\begin{figure}[hptb]
\minipage{0.51\textwidth} \includegraphics[width=2.6in]{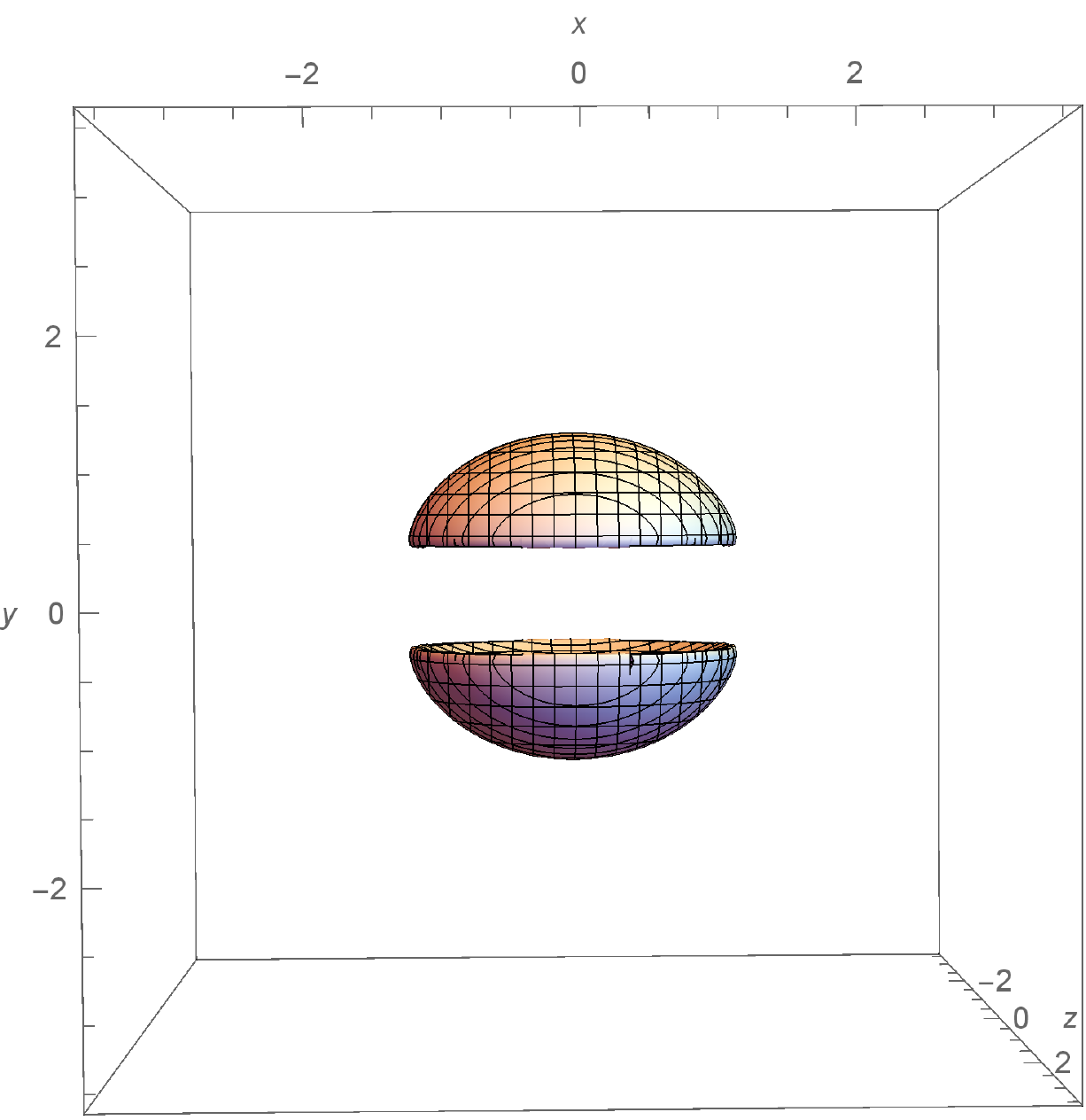}
\endminipage
\minipage{0.51\textwidth} \includegraphics[width=2.6in]{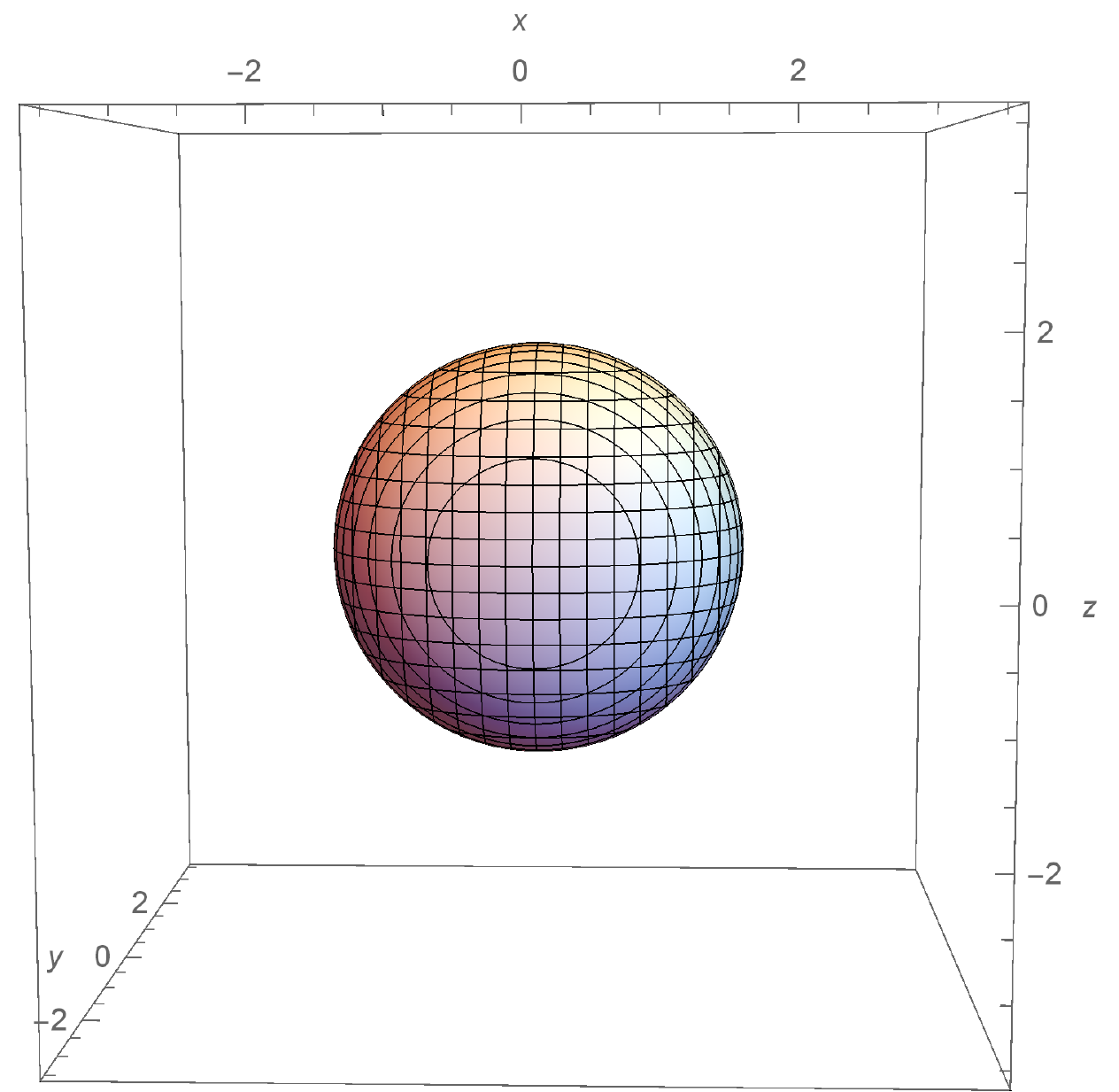}
\endminipage
\caption{The event horizon for the charged non-Kerr accelerating black hole. In the left panel, the parameters are $M=1,\epsilon =0.5,a=0.7,q=0.6$ and $\alpha =0.1$. In the right panel the parameters are $M=1,\epsilon =-0.7,a=0.7$, $q= 0.6$ and $\alpha =0.07.$}
\label{event}
\end{figure}

The equation $g^{rr}=0$,
where $g^{rr}$ is the inverse of the $g_{rr}$ component of the metric, gives the event horizon, which will be denoted by $r_{+}$. For metric (\ref{2.1}), the event horizon equation is
\begin{equation}
 \frac{Q(1-\alpha ^{2}r^{2})+a^2 h\sin^2{\theta}}{\rho^2 \Omega ^2}=0.  \label{2.4}
\end{equation}
For $h=0$, the above equation reduces to the horizon equation for charged accelerating solution. In this case, Eq. (\ref{2.4}) gives the event and Cauchy horizons (which are same as in the Kerr-Newman case) and also acceleration horizons at $r_{A}=1/\alpha$ and $r_{\alpha}=1/\alpha \cos \theta$ \cite{gp}. 
For $h\neq0$, the horizon equation is
\begin{equation}
 \frac{(a^{2}+q^{2}-2Mr+r^{2})(1-\alpha ^{2}r^{2})+a^2 h\sin^2{\theta}}{\rho^2 \Omega ^2}=0,  \label{2.4a}
\end{equation}
which shows that the acceleration horizon $r_{A}=1/\alpha$ does not exist. 
The presence of $h$ introduces the $\theta$-dependence onto the event horizon. Such dependence has been observed for other spacetimes as well \cite{jp,jpc,dh,cm}.
Due to the complexity of the equation, the event horizon equation  is solved numerically.
The numerical plots are shown in Fig. \ref{event}.

\begin{figure}[hptb]
\begin{subfigure}{.3\textwidth}
\centering
\includegraphics[width=5cm,height=6cm]{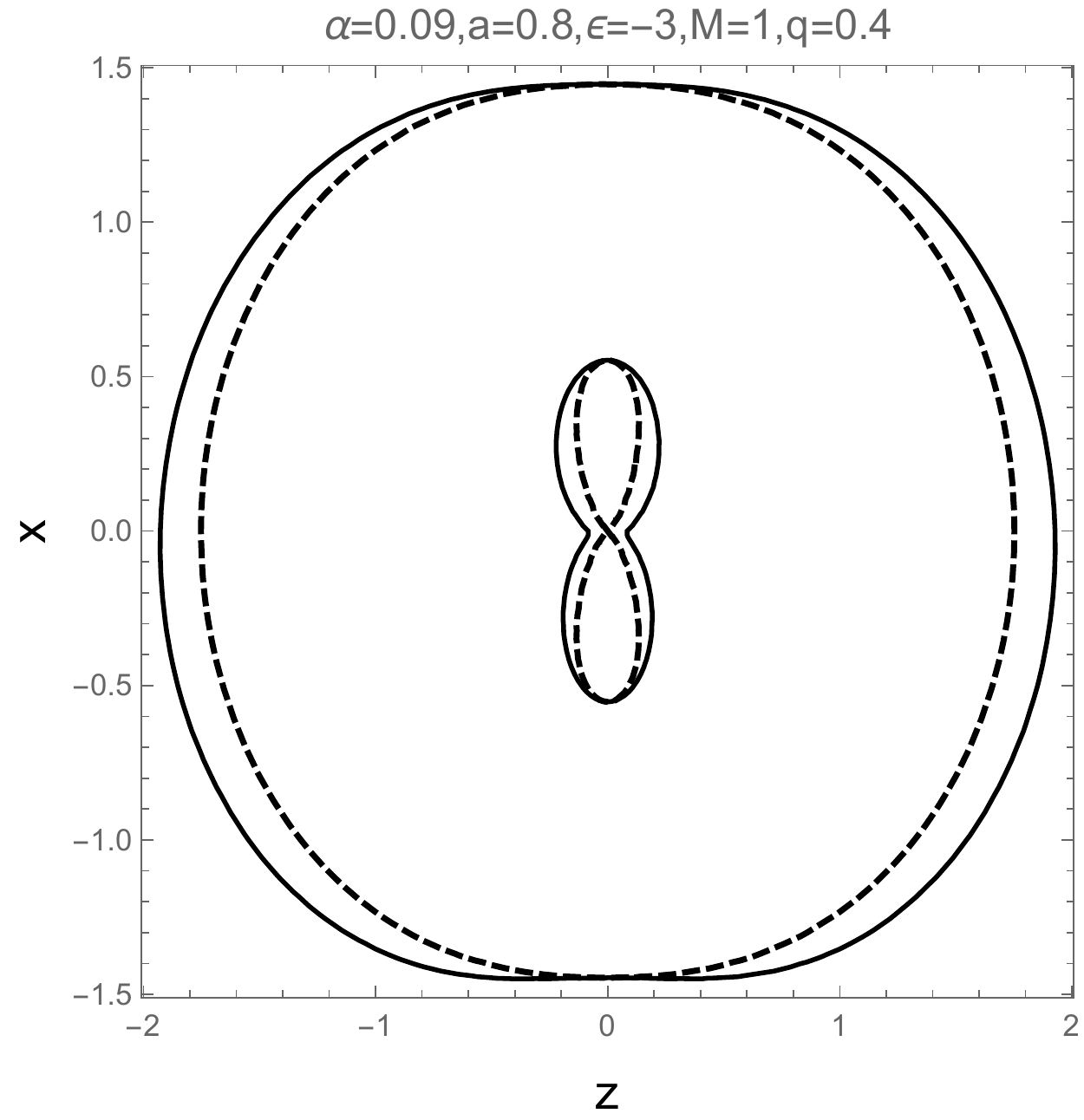}
\end{subfigure}\hfill
\begin{subfigure}{.3\textwidth}
\centering
\includegraphics[width=5cm,height=6cm]{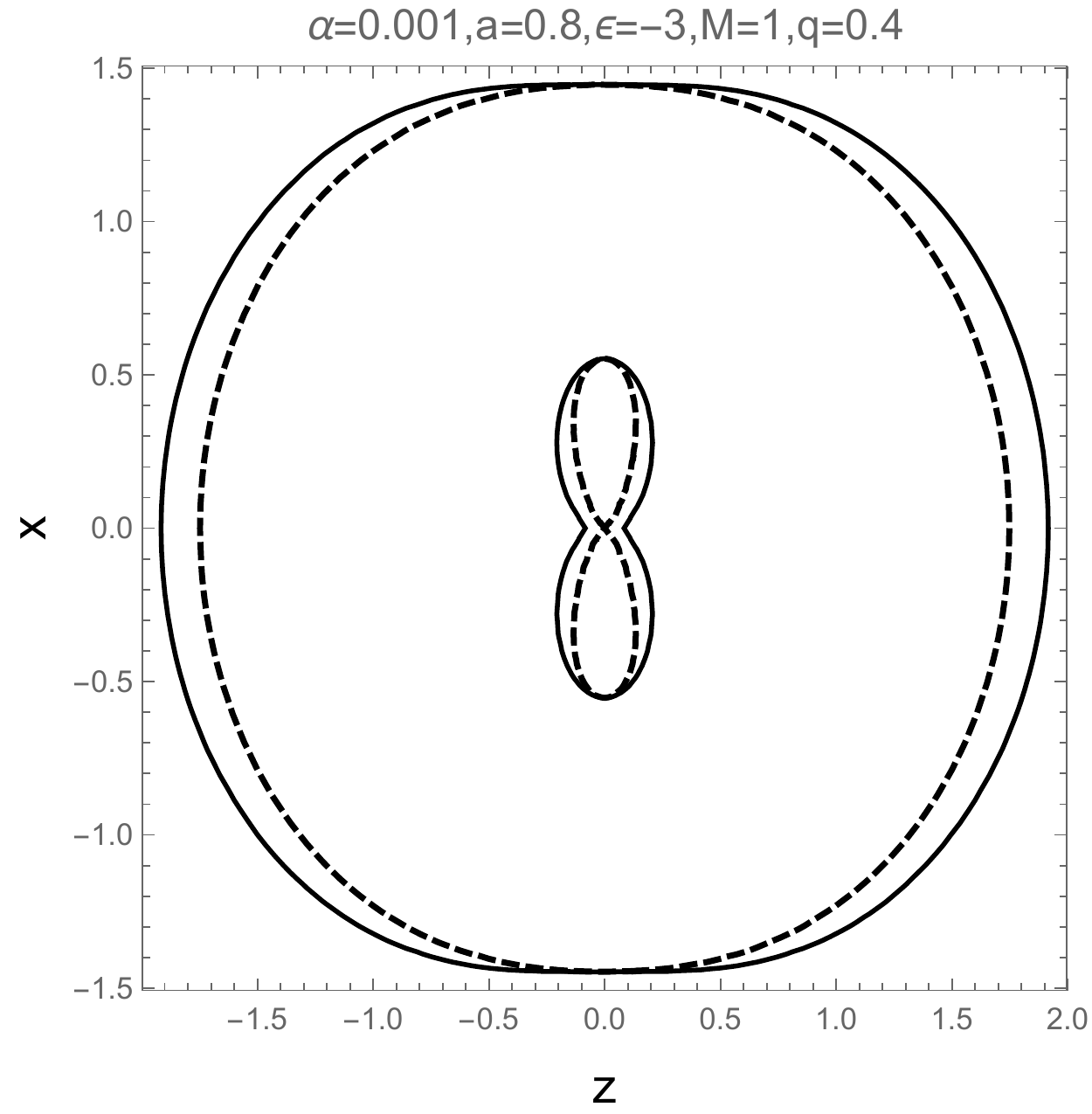}
\end{subfigure}\hfill
\begin{subfigure}{.3\textwidth}
\centering
\includegraphics[width=5cm,height=6cm]{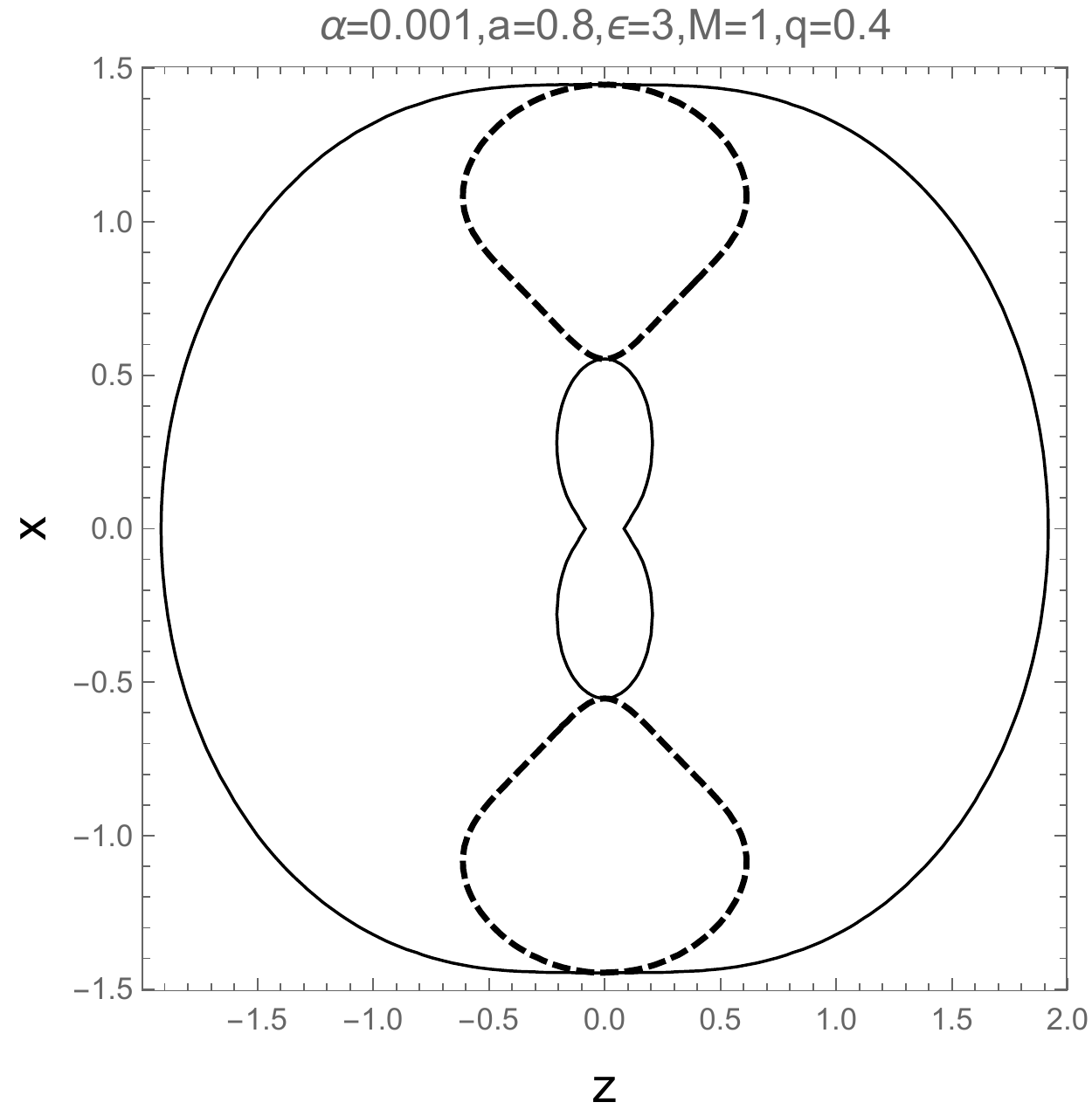}
\end{subfigure}
\caption{The ergosphere for the charged non-Kerr accelerating spacetime.  The solid curve shows the infinite redshift surface and dashed curve represents the event horizon.}\label{chergo}
\end{figure}

From Fig. \ref{event} it is clearly seen that the event horizon exhibits two kinds of behaviour.
In the right panel,  the horizon shows spherical topology while on the left panel, two disjoined horizons are obtained.
Such disjoined event horizon has also been observed for some other modified Kerr black holes as well \cite{dh, dh1}.
The metric has the determinant
\begin{align}
det(g_{\mu\nu})=&-\frac{(h+1)^2 \sin ^{2} \theta}{\Omega ^8 (a^2 h \sin ^{2} \theta+Q)}
\Bigg[ a^4 (h+1) Q \sin ^{4} \theta+a^2 (a^2+r^2) \sin ^{2} \theta (a^2 h P+h P r^2
\nonumber
\\
&
-2 (h+1) Q)
+Q (a^2+r^2)^2\Bigg].
\label{det1}
\end{align}
The Kretschmann scalar for metric (\ref{2.1}) is a long and complex expression which shows divergent behavior at $1+h=0$. Also, from Eq. (\ref{det1}), it is evident that the determinant vanishes at $1+h=0$. Therefore, the metric has an intrinsic singularity at the surface given by $1+h=0.$ This is in contrast to the accelerating black hole metric which has a Kerr-like singularity at $r=0,\theta=\pi/2$.
The infinite redshift surface is obtained by solving the equation
\begin{equation}
g_{tt}=0,
\end{equation}
 which for the black hole metric of Eq. (\ref{2.1}) is
\begin{equation}
[(a^{2}+q^{2}-2Mr+r^{2})(1-\alpha ^{2}r^{2})-a^{2}(1-2\alpha M\cos \theta +\alpha ^{2}\left(q^{2}+a^{2}\right)\cos ^{2}\theta )\sin^{2}\theta](1+h)=0.
\end{equation}
It is clear from this equation that there are two candidates for the infinite redshift surface, $1+h=0$ being one of them. But $1+h=0$ is the singularity, so it is concluded that infinite redshift surface exists at the solution of $Q-a^2 P\sin^{2}\theta=0$. Thus the infinite redshift surface is same as in the charged accelerating black hole metric. The region between the outer infinite redshift surface and the event horizon is known as the ergosphere and it is plotted in Fig. \ref{chergo}.

\begin{figure}[hptb]
\begin{subfigure}{.3\textwidth}
\centering
\includegraphics[width=5cm,height=6cm]{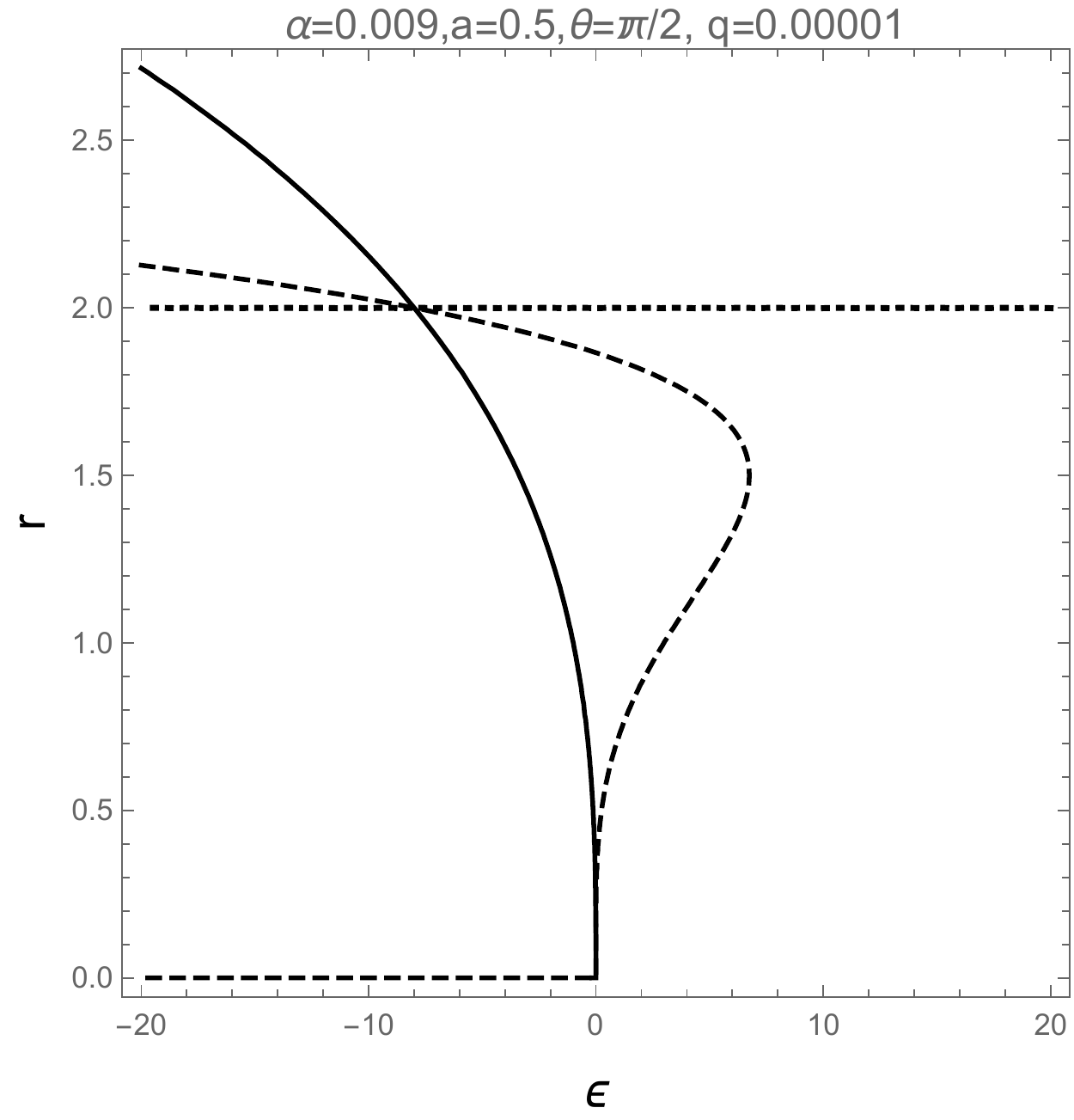}
\end{subfigure}\hfill
\begin{subfigure}{.3\textwidth}
\centering
\includegraphics[width=5cm,height=6cm]{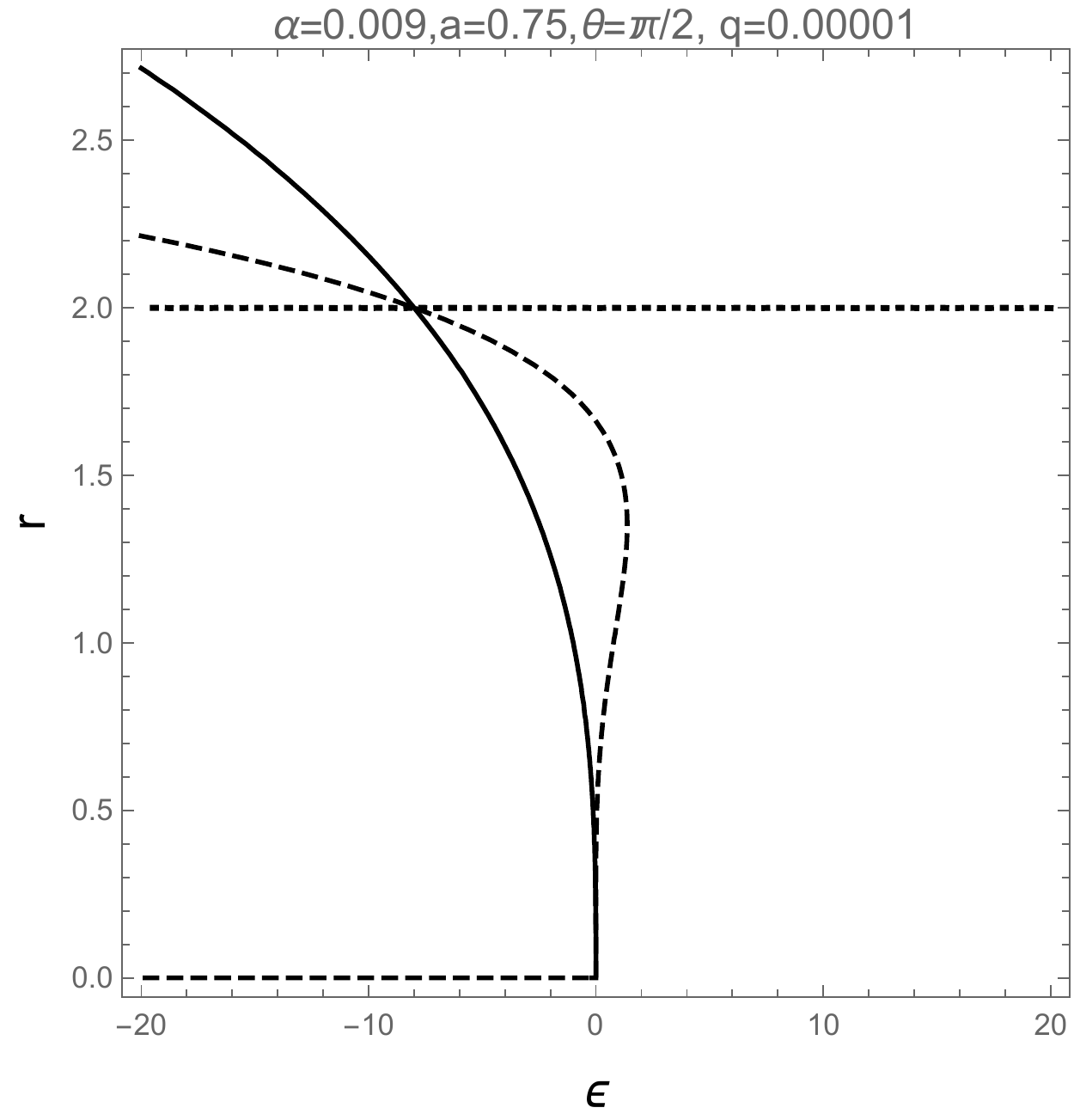}
\end{subfigure}\hfill
\begin{subfigure}{.3\textwidth}
\centering
\includegraphics[width=5cm,height=6cm]{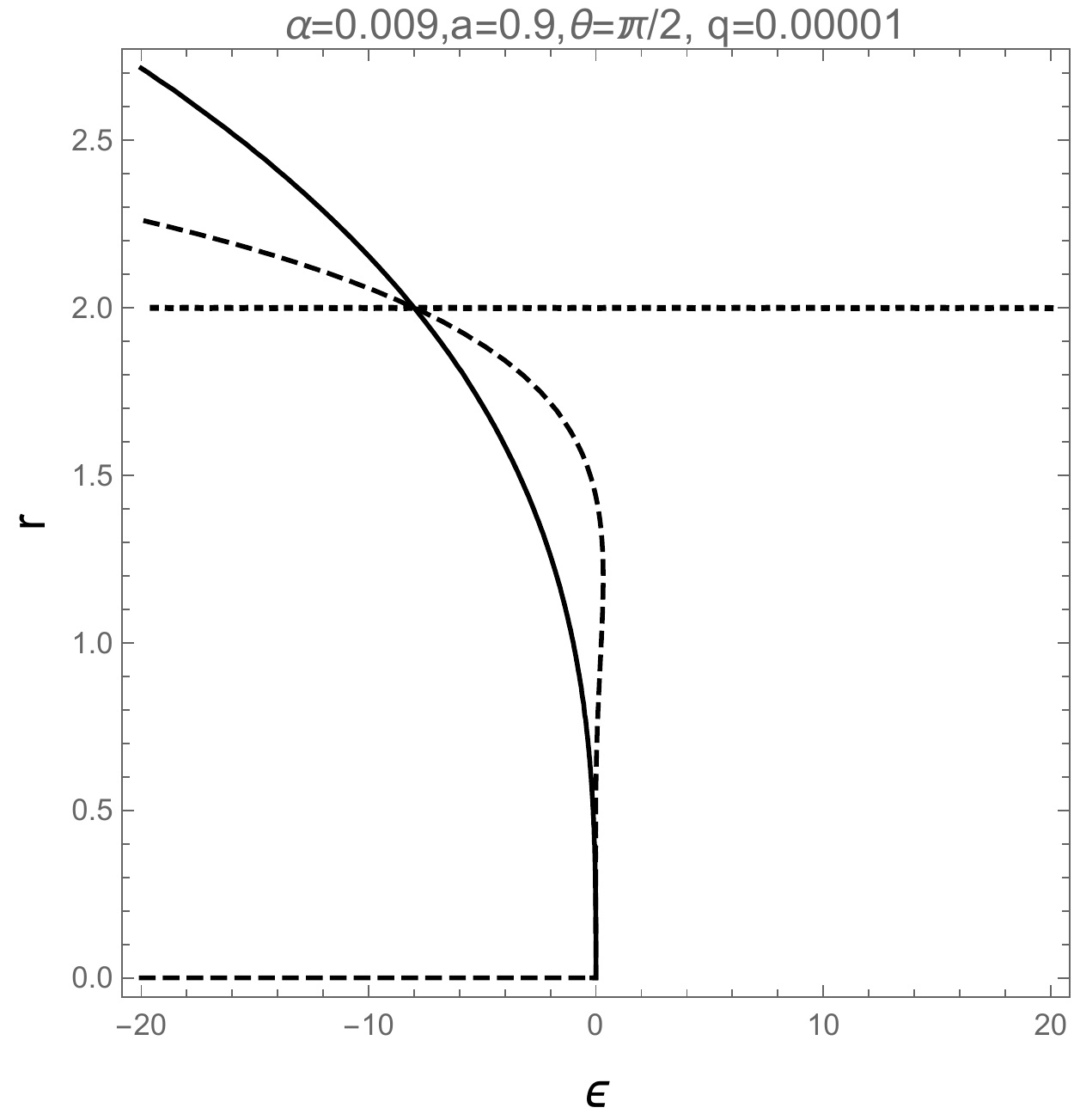}
\end{subfigure}
\caption{The charged non-Kerr accelerating black hole. The graphical representation of the event horizon (dashed curve), the infinite red shift surface (dotted line) and $1+h=0$ (the solid curve) for varying values of the rotation parameter, $a=0.5$, $a=0.75$ and $a=0.9$. We have taken the acceleration parameter as $0.009$, $\theta=\pi/2$ and $q=0.00001.$ Mass has been set to 1 in these plots. } \label{redshift}
\end{figure}

The graphical representation of the event horizon, the infinite red shift surface and $1+h=0$ for varying numerical values of parameters are plotted in Fig. \ref{redshift}.
\subsection{The uncharged case}
\begin{figure}[hptb]
\minipage{0.51\textwidth} \includegraphics[width=2.6in]{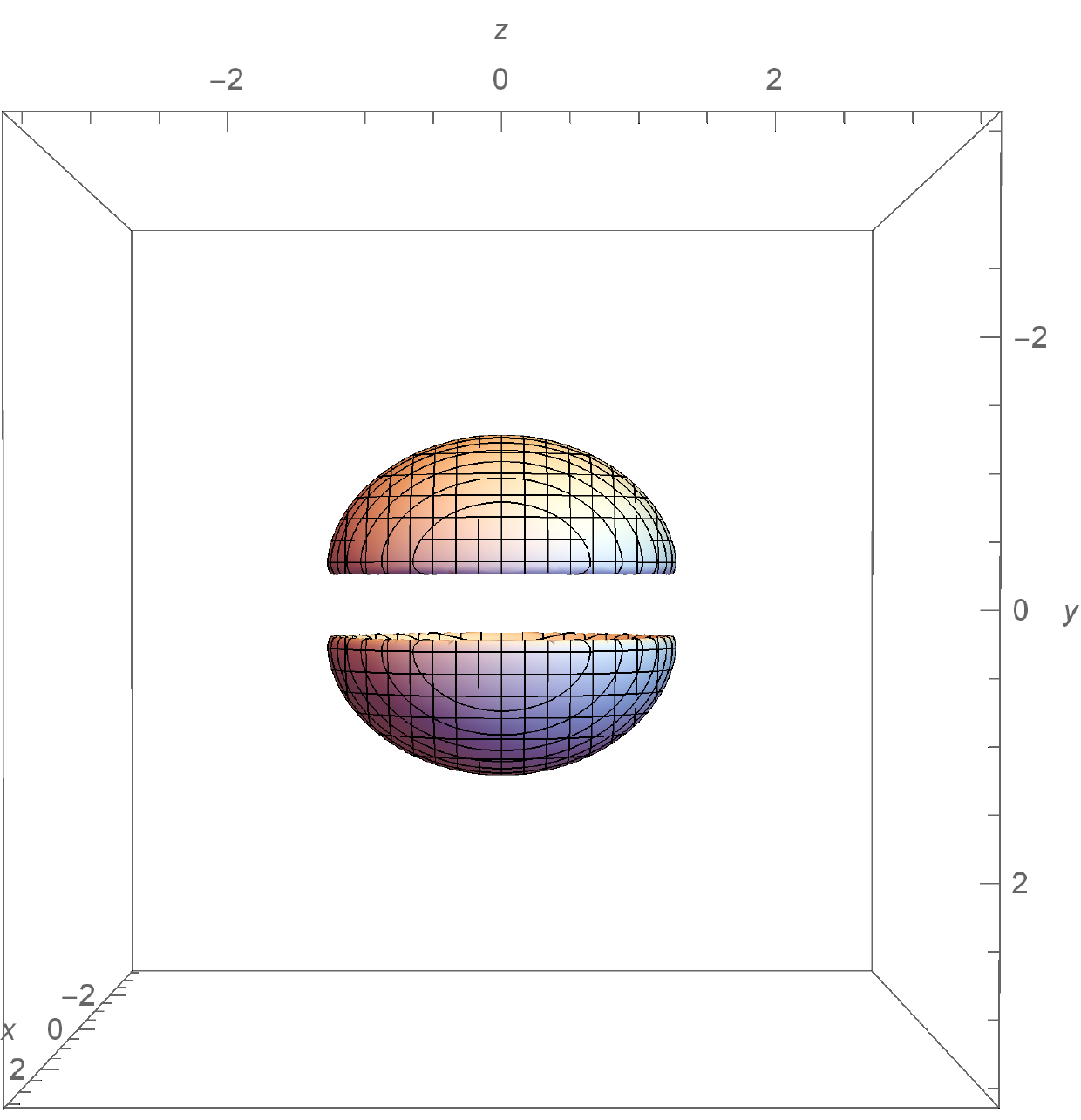}
\endminipage
\minipage{0.51\textwidth} \includegraphics[width=2.6in]{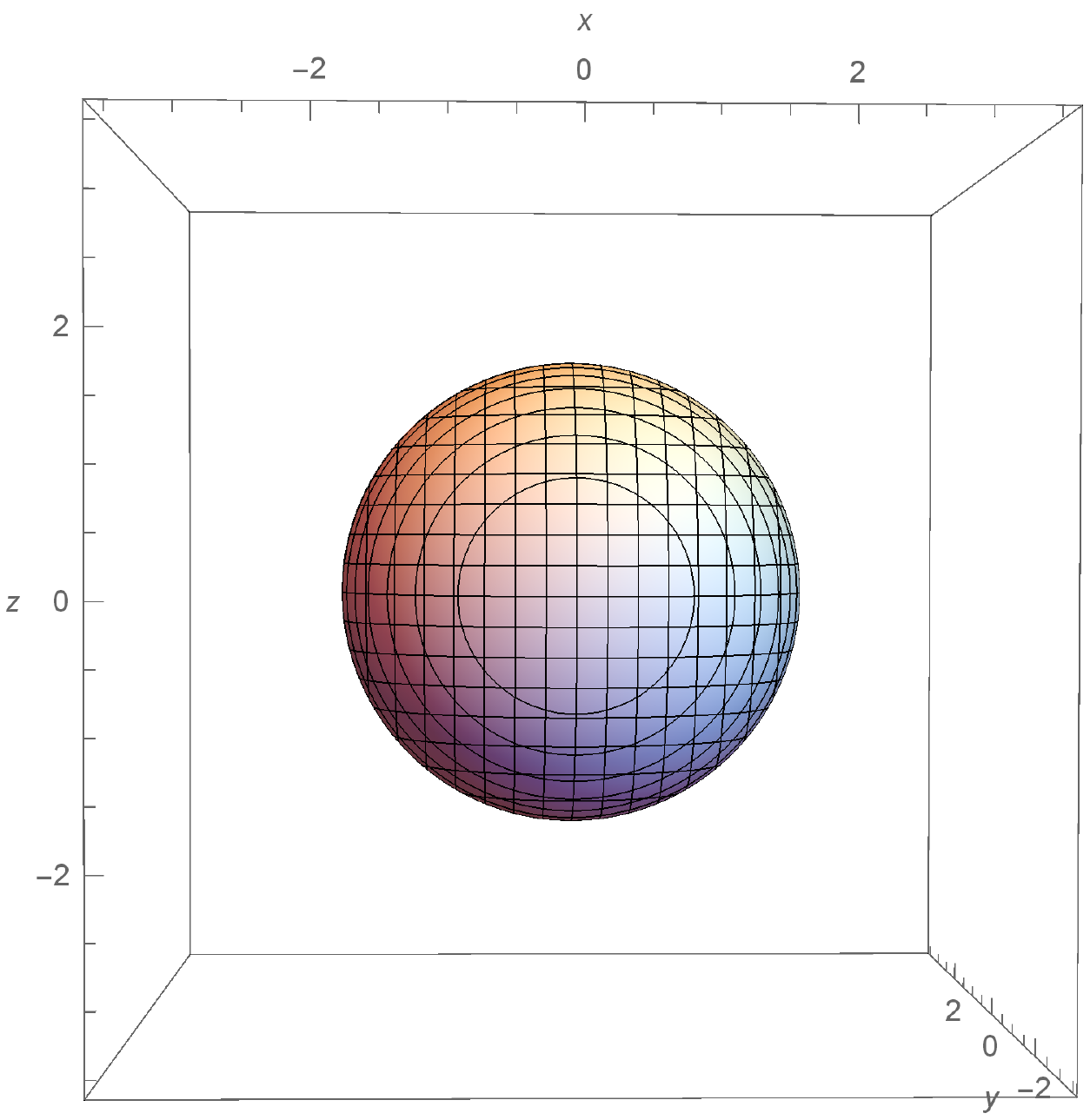}
\endminipage
\caption{The event horizon for the non-Kerr accelerating spacetime. In the left panel we have taken $M=1,\epsilon =0.4,a=0.89$ and $\alpha =0.1$. In the right panel the parameters are $M=1,\epsilon =0.04,a=0.4$ and $\alpha =0.06.$}
\label{event2}
\end{figure}

If we put $q=0$ in spacetime (\ref{2.1}), we get the non-Kerr accelerating black hole spacetime. It is given by
\begin{align}
ds^{2} &=\frac{1}{\Omega ^{2}}\Big\{-\big(\frac{Q}{\rho ^{2}}-\frac{a^{2}P\sin
^{2}\theta }{\rho ^{2}}\big)\left(1+h\right)dt^{2}+\frac{\rho ^{2}\left(1+h\right)}{Q+a^2h\sin^{2} \theta}dr^{2}+\frac{\rho ^{2}}{P}%
d\theta ^{2}  \nonumber \\
&{+}\sin ^{2}\theta\big(\frac{P(r^{2}+a^{2})^{2} }{\rho ^{2}}{-}\frac{Qa^{2}\sin
^{2}\theta \left(1+h\right)}{\rho ^{2}}\big)\Big\} d\phi ^{2}{-}\frac{2a\sin ^{2}\theta
\left[P\left(r^{2}+a^{2}\right)-Q\right]\left(1+h\right) }{\rho ^{2}\Omega ^{2}}dtd\phi, \label{2.5}
\end{align}
where $\Omega$ and $\rho^2$ are same as in Section 1 and $P$ and $Q$ have the expressions given as
\begin{align}
&P =1-2\alpha M\cos \theta +\alpha ^{2}a^{2}\cos ^{2}\theta , \label{2.5a} \\&
Q=(a^{2}-2Mr+r^{2})(1-\alpha ^{2}r^{2}).
\label{2.7}
\end{align}
As with the charged case, it is assumed here, that the above metric is a vacuum
solution to the unknown field equations which are different from Einstein's field equations for non-zero $h(r,\theta)$.
The numerical plots of the event horizon for this black hole are shown in the Fig. \ref{event2}. As in the charged case, we again observe two kinds of behavior for the event horizon.

\begin{figure}[H]
\begin{subfigure}{.3\textwidth}
\centering
\includegraphics[width=5cm,height=6cm]{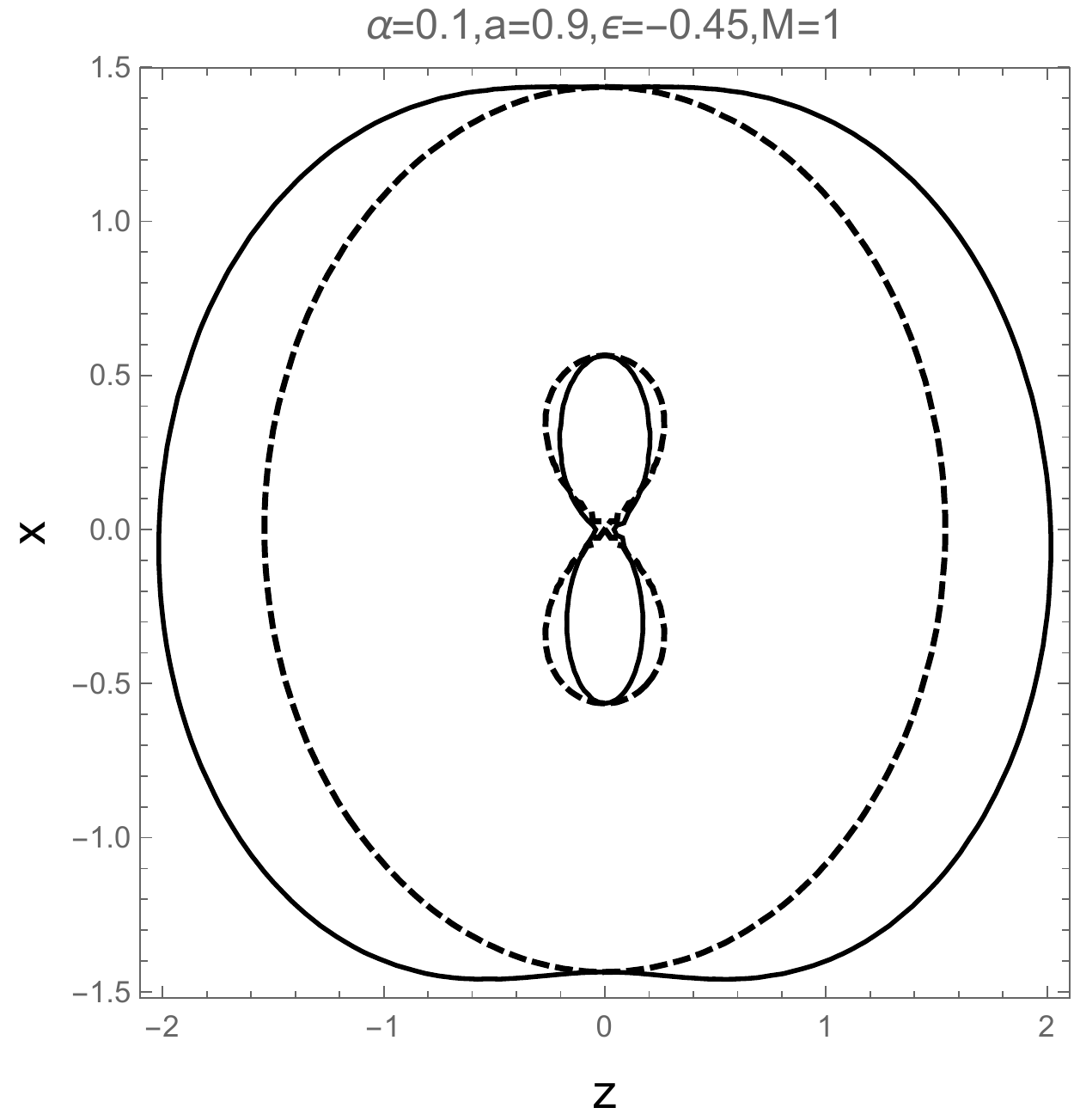}
\end{subfigure}\hfill
\begin{subfigure}{.3\textwidth}
\centering
\includegraphics[width=5cm,height=6cm]{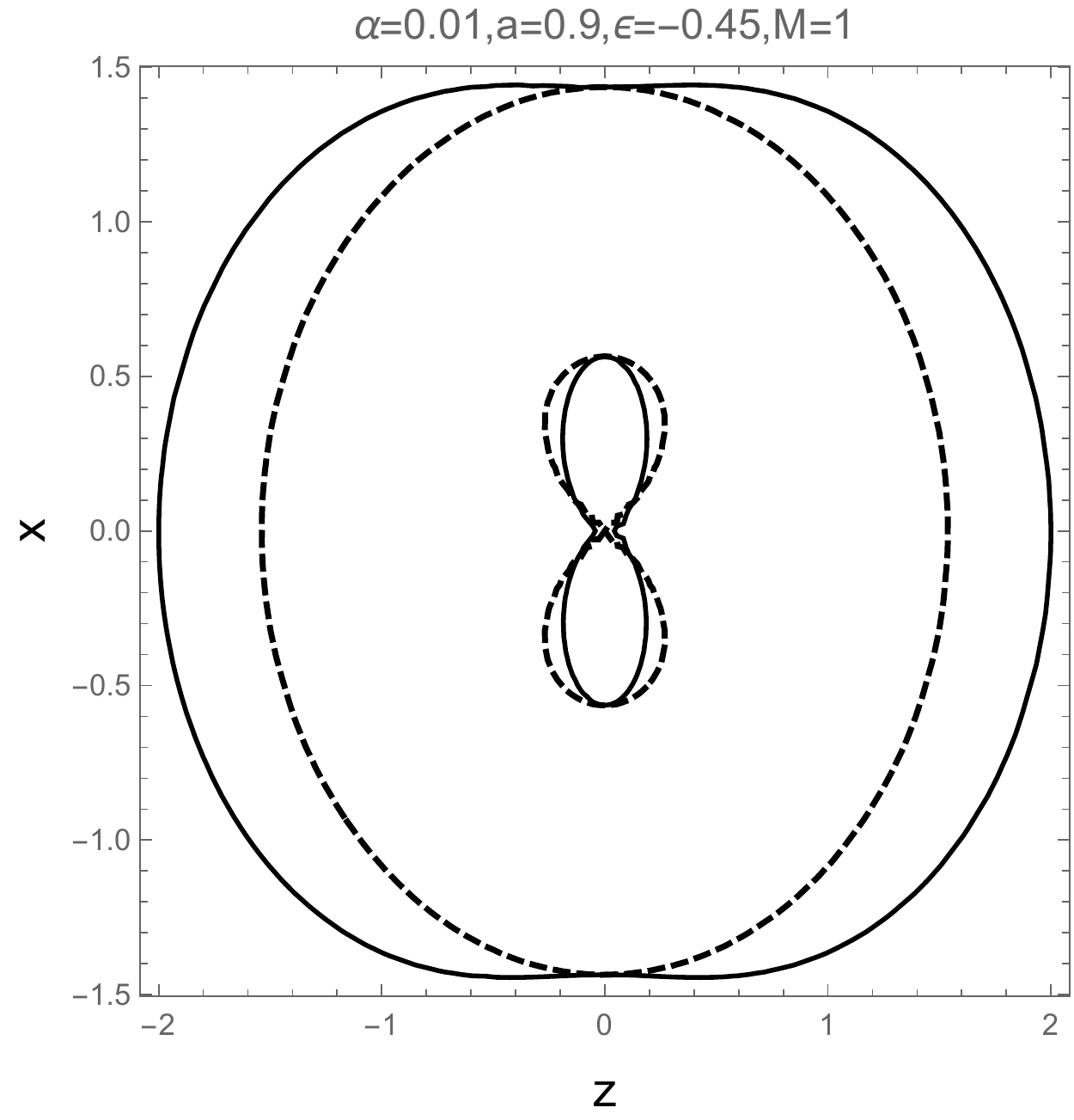}
\end{subfigure}\hfill
\begin{subfigure}{.3\textwidth}
\centering
\includegraphics[width=5cm,height=6cm]{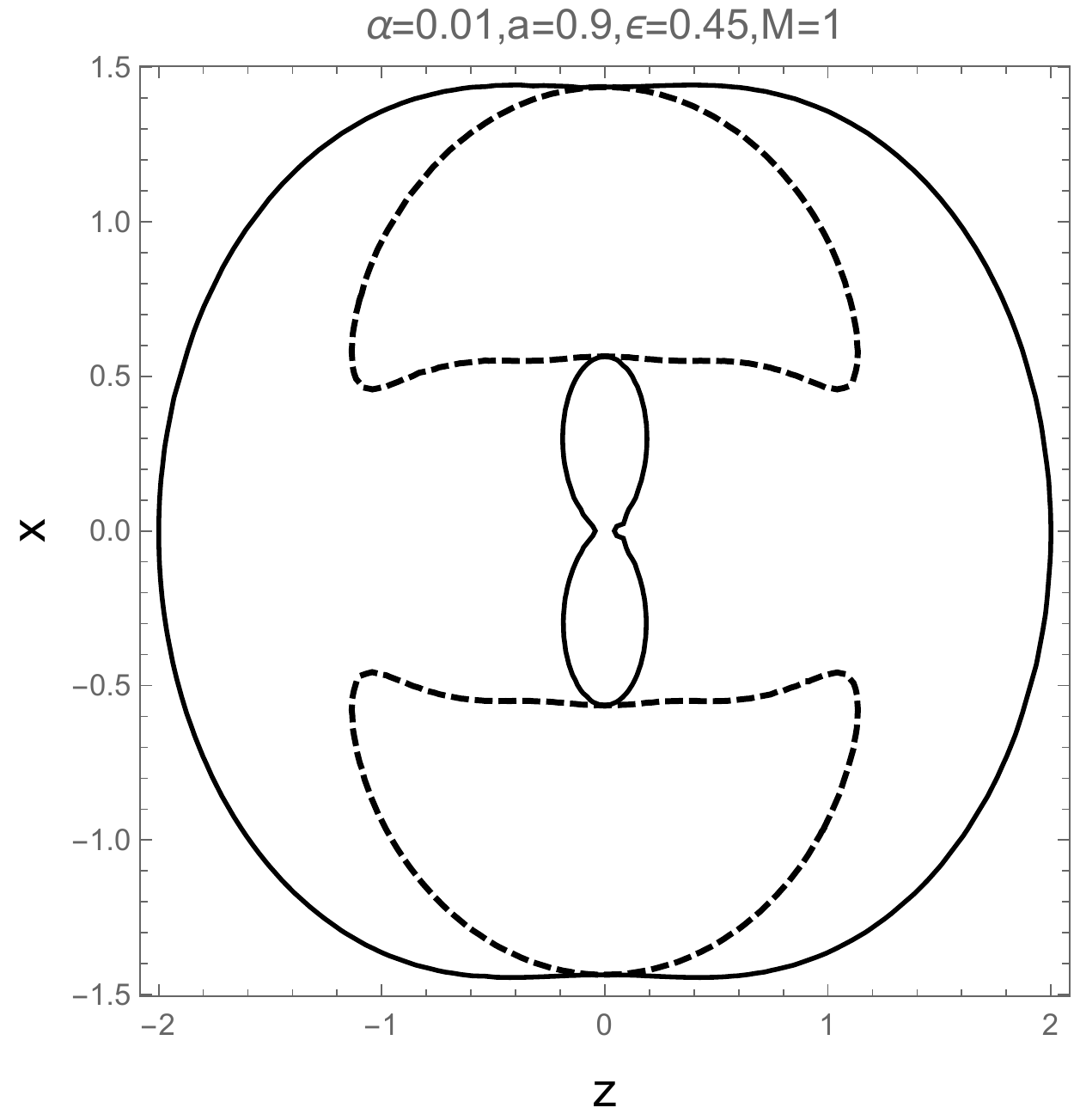}
\end{subfigure}
\caption{The ergosphere of the non-Kerr accelerating black hole for various values of the parameters. Here the solid curve is the infinite redshift surface and the dashed curve denotes the event horizon.}\label{nergo}
\end{figure}

The spacetime, in this case, also has an intrinsic singularity at $1+h=0$. 
The $g_{tt}=0$ gives
\begin{equation}
(a^{2}-2Mr+r^{2})(1-\alpha ^{2}r^{2})-a^{2}(1-2\alpha M\cos \theta +a^2 \alpha ^{2}\cos ^{2}\theta )\sin^{2}\theta=0,
\end{equation}
showing the $\epsilon$ independent behavior. The ergosphere is shown in Fig. \ref{nergo}.


\begin{figure}[H]
\begin{subfigure}{.3\textwidth}
\centering
\includegraphics[width=5cm,height=6cm]{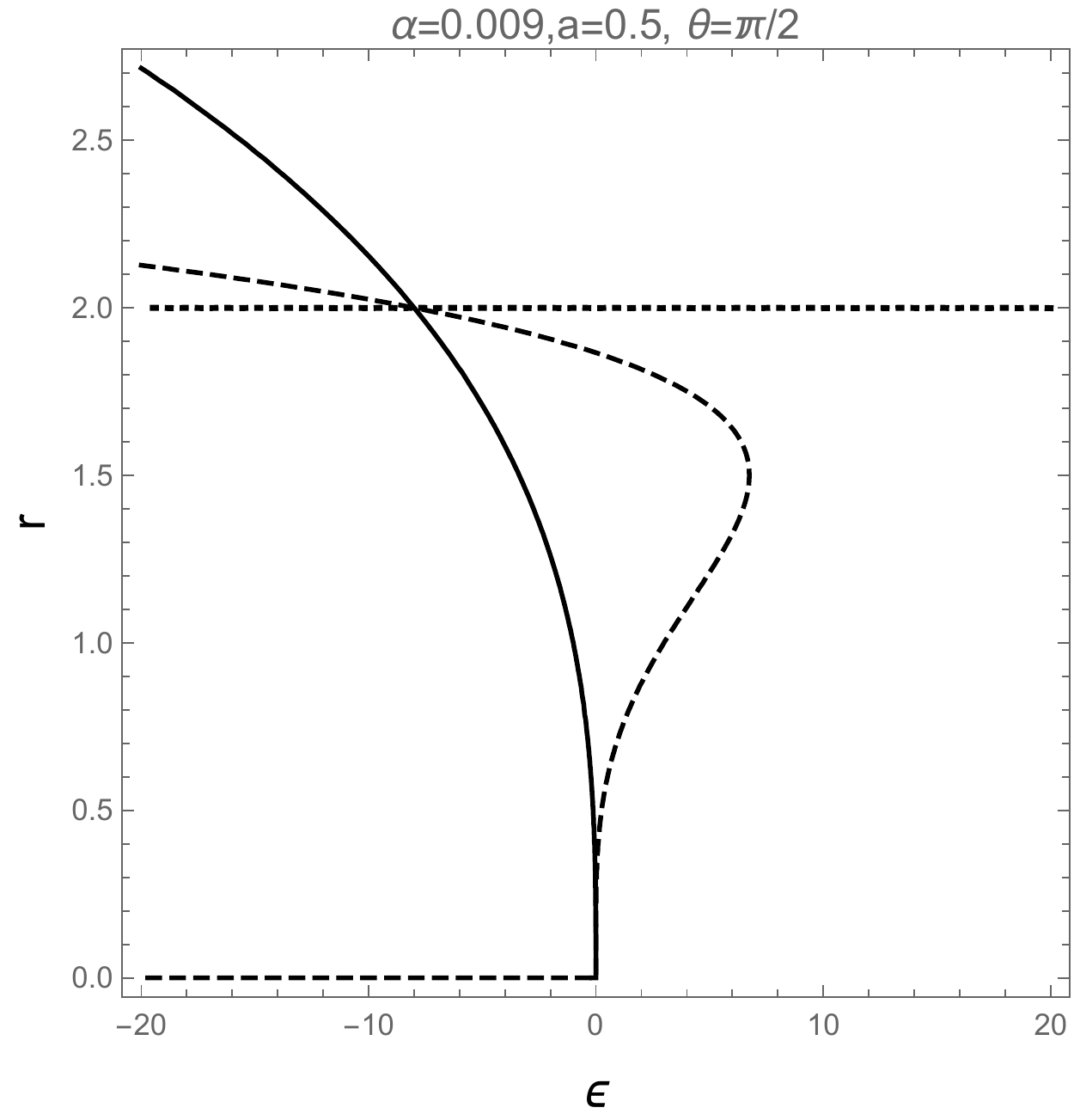}
\end{subfigure}\hfill
\begin{subfigure}{.3\textwidth}
\centering
\includegraphics[width=5cm,height=6cm]{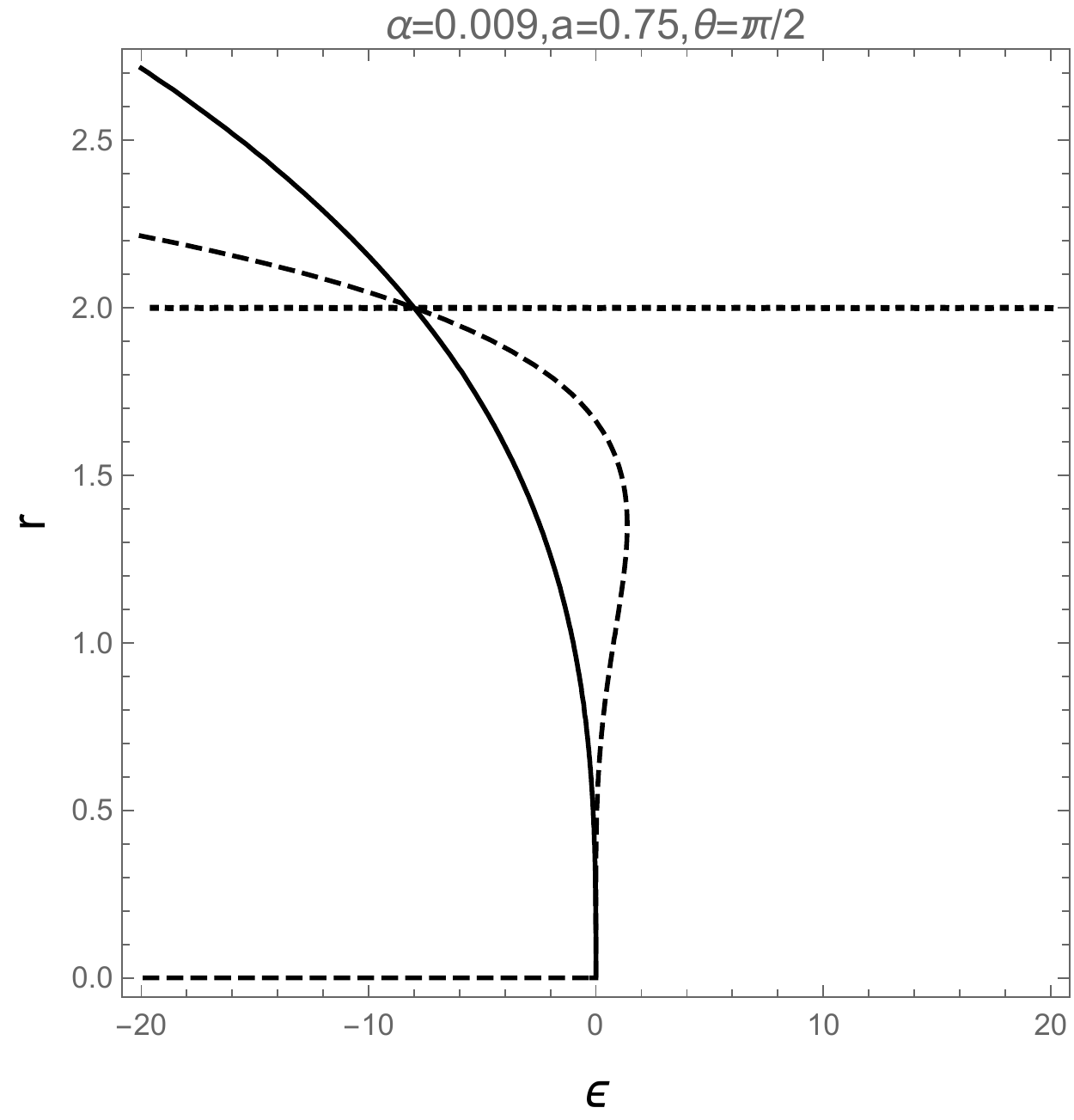}
\end{subfigure}\hfill
\begin{subfigure}{.3\textwidth}
\centering
\includegraphics[width=5cm,height=6cm]{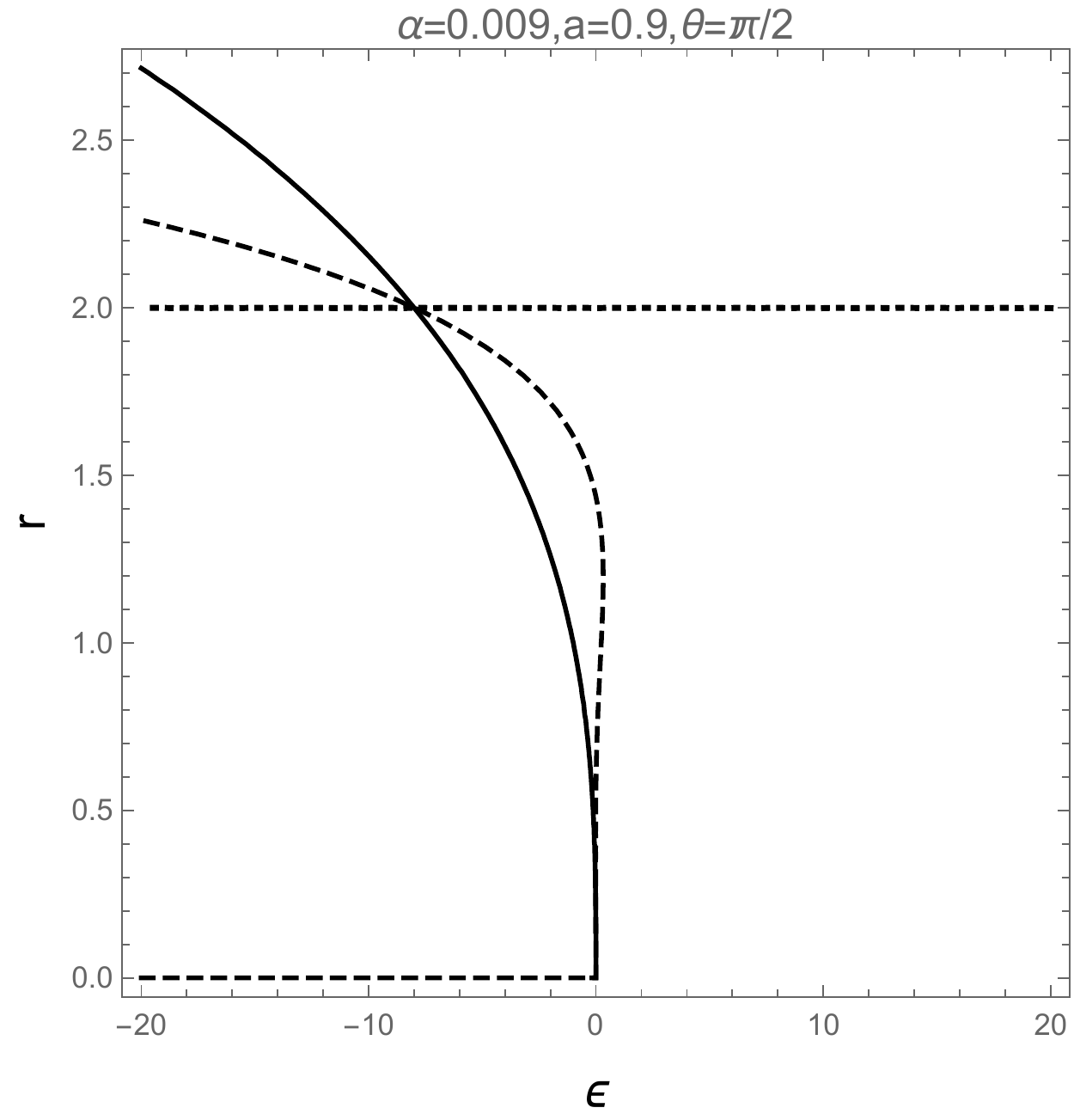}
\end{subfigure}
\caption{The uncharged non-Kerr accelerating black hole. The graphical representation of the event horizon (dashed curve), the infinite red shift surface (dotted line) and $1+h=0$ (the solid curve) for varying values of the rotation parameter, $a=0.5$, $a=0.75$ and $a=0.9$, with $\alpha=0.009$, $M=1$ and $\theta=\pi/2$.}\label{redshift1}
\end{figure}

The graphical representation of the event horizon, the infinite red shift surface and $1+h=0$ for varying numerical values of the parameters are plotted in Fig. \ref{redshift1}.

\section{Thermodynamics of the non-Kerr accelerating spacetime}\label{thermo2}
 This section comprises the thermodynamic analysis of the two
 black hole metrics presented in the previous section.
\subsection{The charged case}
 The study of thermodynamic features is an important property of a black hole. To satisfy the first law of thermodynamics for black holes,  we need to find the Hawking temperature that becomes proportional to surface gravity near the horizon and an entropy which is proportional to the area of horizon. The surface gravity $\kappa$ is given as 
\begin{equation}
\kappa=\frac{1}{\sqrt{-h}}\frac{\partial}{\partial x^a}\left(\sqrt{-h}h^{ab}\frac{\partial}{\partial x^b}\right), \label{4.1}
\end{equation}
where $h^{ab}$ denotes the inverse of the metric deduced from the $t-r$ sector of the spacetime
and $h = \det h_{ab}$. Since the metric is stationary, so $h^{00}=0$. Putting the values of $h^{11}$ and $\sqrt{-h}$ from Eq.  (\ref{2.1}), the surface gravity takes the form
\begin{equation}
\kappa{=}\frac{\Omega^{2}\sqrt{(a^{2}{+}q^{2}{-}2Mr{+}r^{2})(1{-}\alpha ^{2}r^{2}){+}a^2h\sin^{2}\theta}}{(1{+}h)\sqrt{(a^{2}{+}q^{2}{-}2Mr{+}r^{2})(1{-}\alpha ^{2}r^{2}){-}a^2(1{-}2\alpha M\cos \theta {+}\alpha ^{2}\left(a^{2}{+}q^{2}\right)\cos ^{2}\theta)\sin^{2}\theta}}\frac{\partial W(r,\theta)}{\partial r},
\end{equation}
where
\begin{eqnarray}
W(r,\theta)&=&\frac{1}{\rho^2}\Bigg[\left[(a^{2}{+}q^{2}{-}2Mr{+}r^{2})(1{-}\alpha ^{2}r^{2}){+}a^2h\sin^{2}\theta\right]
\nonumber \\ 
&\times &
\left[(a^{2}{+}q^{2}{-}2Mr{+}r^{2})(1{-}\alpha ^{2}r^{2}){-}a^2(1{-}2\alpha M\cos \theta {+}\alpha ^{2}\left(a^{2}{+}q^{2}\right)\cos^{2}\theta \right]\sin^{2}\theta \Bigg]^{1/2}.
\end{eqnarray}
 At the event horizon $r_{+}$, the surface gravity $\kappa_{+}$ takes the form
\begin{eqnarray}
\kappa_{+}&=&\frac{\Omega^{2}}{2(1+h)\rho^2}\Bigg\{(1-\alpha ^{2}r_{+}^{2})(2r_{+}-2M)-2\alpha^2 r_+ (a^{2}+q^{2}+r_{+}^2-2Mr_{+})\nonumber \\
&&+a^2\epsilon M^3 \sin^2\theta\left[\frac{1}{(r_{+}^{2}+a^2 \cos^2\theta)^2}-\frac{4r_{+}^{2}}{(r_{+}^{2}+a^2 \cos^2\theta)^3}\right]\Bigg\}. \label{4.2}
\end{eqnarray}%
The Hawking temperature, $T=\kappa/2\pi$, at the event horizon can be obtained by substituting the value of $\kappa$ from the above equation as 
\begin{eqnarray}
T&=&\frac{\Omega^{2}}{4\pi(1+h)\rho^2}\Bigg\{(1-\alpha ^{2}r_{+}^{2})(2r_{+}-2M)-2\alpha^2 r_+ (a^{2}+q^{2}+r_{+}^2-2Mr_{+})\nonumber \\
&&+a^2\epsilon M^3 \sin^2\theta\left[\frac{1}{(r_{+}^{2}+a^2 \cos^2\theta)^2}-\frac{4r_{+}^{2}}{(r_{+}^{2}+a^2 \cos^2\theta)^3}\right]\Bigg\}.
\end{eqnarray}
\begin{figure}[H]
\centering
\includegraphics[width=3in]{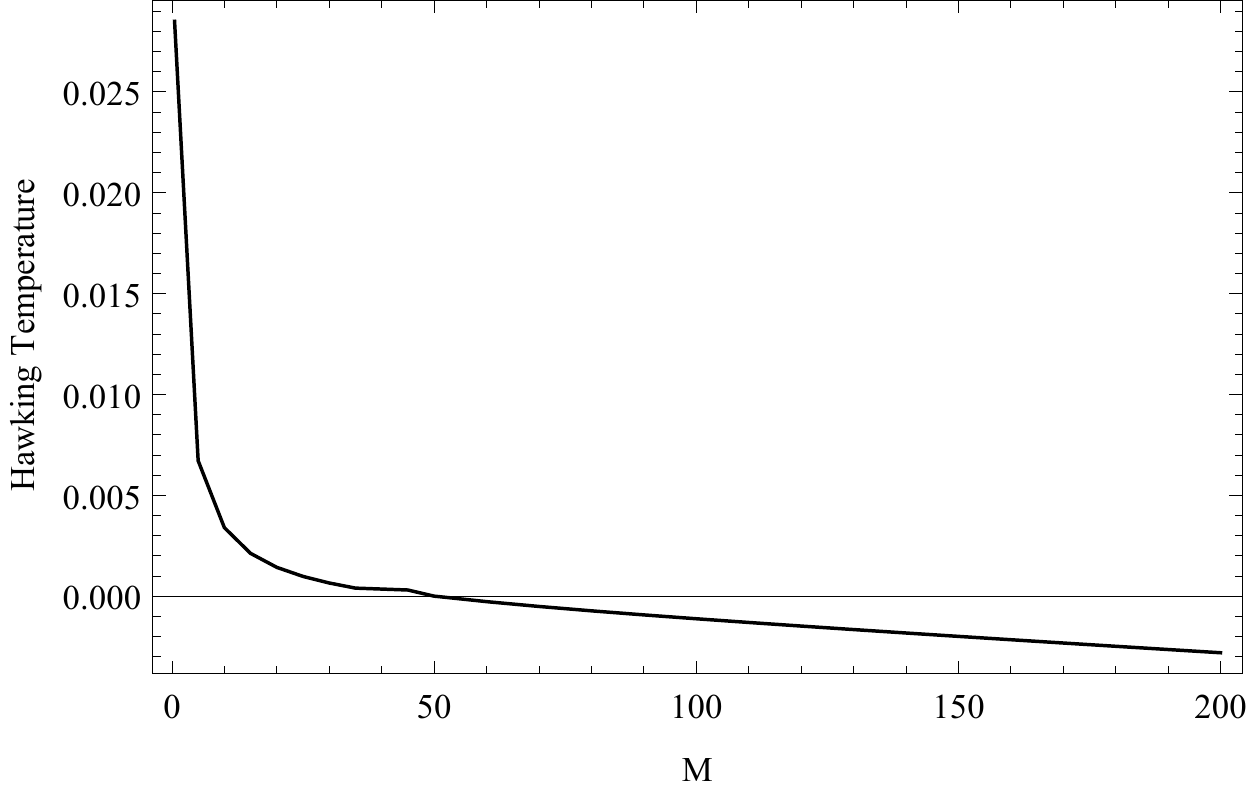}
\caption{Plot of the Hawking Temperature against mass for the charged non-Kerr accelerating black hole for $a=0.5$, $\epsilon=0.5$, $\theta=\pi/2$, $\alpha=0.01$ and $q=0.5.$}\label{cht}
\end{figure}
The Hawking temperature at the equatorial plane has been plotted in Fig. \ref{cht}. The plot shows that temperature exhibits negative behavior for various mass values. Such negative temperature has also been observed in the case of accelerating black holes \cite{kb}.

\begin{figure}[H]
\centering
\includegraphics[width=3in]{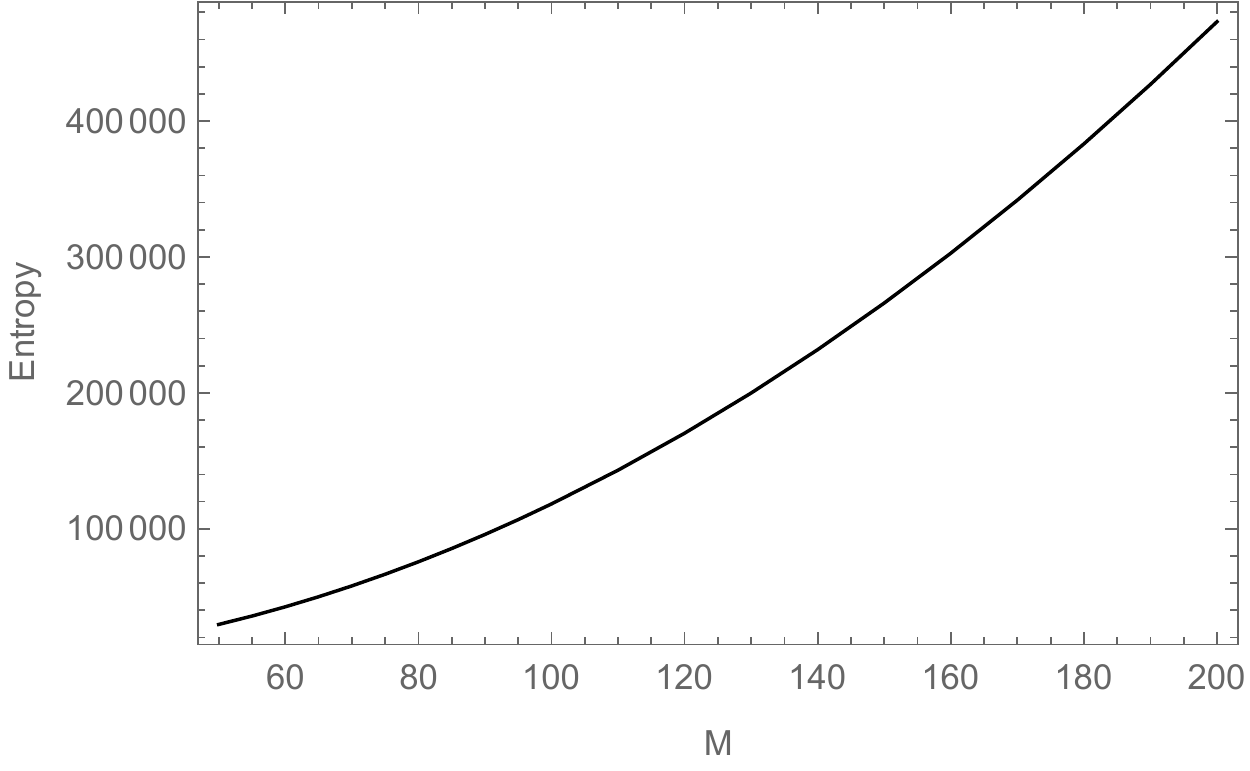}
\caption{Plot of the entropy against mass for the charged non-Kerr accelerating black hole for $a=0.5$, $\epsilon=0.5$, $\theta=\pi/2$, $\alpha=0.01$ and $q=0.5$.}
\label{cent}
\end{figure}

To discuss entropy of the black holes, the first thing to be determined is the angular velocity $\omega$, which leads to the horizon area \cite{av}. We have 
\begin{equation}
\omega=\frac{\partial \phi}{\partial t}=-\frac{g_{t\phi}}{g_{\phi\phi}}.
\end{equation}
Substituting the required metric components, the angular velocity at the event horizon is given as
\begin{equation}
\omega_{+}=\frac{aP\left[(r_{+}^2+a^2)-Q\right]\left(1+h\right)}{P\left(r_{+}^2+a^2\right)^2-Qa^2 \sin^2\theta(1+h)}. \label{4.3}
\end{equation}
The horizon area of a rotating black hole is defined as 
\begin{equation}
A=\frac{4\pi a}{\omega_{+}}.
\end{equation}
Now we use Eq. (\ref{4.3}) to obtain
\begin{equation}
A=\frac{4\pi\left[ P\left(r_{+}^2+a^2\right)^2-Qa^2 \sin^2\theta(1+h)\right] }{P\left[(r_{+}^2+a^2)-Q\right]\left(1+h\right)}.\label{area}
\end{equation}
The entropy of the black hole, $S=A/4$, by substituting from the above equation, takes the form    

\begin{equation}
S=\frac{\pi \left[P\left(r_{+}^2+a^2\right)^2-Qa^2 \sin^2\theta(1+h)\right]}{\left[P(r_{+}^2+a^2)-Q\right]\left(1+h\right)}.
\end{equation}
The above entropy expression has dependence on the $r_{+}$ and $\theta$. For the graphical purpose, the entropy has been plotted at the equitorial plane ($\theta=\pi/2$) in Fig. \ref{cent}. The figure clearly shows that $S$ increases with $M$ in accordance with the second law of black hole thermodynamics.

Now, the first law of thermodynamics can be written in the form of the law of conservation of mass by the following equation \cite{ak, ak1}
\begin{equation}
dm=\frac{\kappa_{+}}{8\pi} dA+ \omega_{+} dJ+\varphi_{+} dq,  \label{4.4}
\end{equation}
where $\varphi_+$ denotes the electrostatic potential of the black hole
\begin{equation}
\varphi_{+}=\frac{4\pi qr_{+}}{A}.
\end{equation}
Substituting the area expression in the above equation, we get
\begin{equation}
\varphi_{+}=\frac{qr_{+}\left[(r_{+}^2+a^2)-Q\right]\left(1+h\right)}{\left(r_{+}^2+a^2\right)^2-Qa^2 \sin^2\theta(1+h)}. \label{4.5}
\end{equation}
By substituting the values of $\kappa_+$, $\omega_+$ and $\varphi_+$ from Eqs. (\ref{4.2}), (\ref{4.3}) and (\ref{4.5}) in
(\ref{4.4}), the first law takes the form
\begin{eqnarray}
dm&=&\frac{\Omega^{2}}{16\pi(1+h)\rho^2}\Bigg\{(1-\alpha ^{2}r_{+}^{2})(2r_{+}-2M)-2\alpha r_+ (a^2+q^{2}+r_{+}^2-2Mr_{+}) \nonumber \\ 
&&+a^2\epsilon M \sin^2\theta 
\left[\frac{1}{(r_{+}^{2}+a^2 \cos^2\theta)^2}-\frac{4r_{+}^{2}}{(r_{+}^{2}+a^2 \cos^2\theta)^3}\right]\Bigg\} dA  \nonumber \\ 
&&+\frac{aP\left[(r_{+}^2+a^2)-Q\right]\left(1+h\right)dJ}{P\left(r_{+}^2+a^2\right)^2-Qa^2 \sin^2\theta(1+h)} 
+\frac{qr_{+}\left[(r_{+}^2+a^2)-Q\right]\left(1+h\right) dq}{\left(r_{+}^2+a^2\right)^2-Qa^2 \sin^2\theta(1+h)}.\label{l1}
\end{eqnarray}
\subsection{The uncharged case }\label{thermo1}
\begin{figure}[hptb]
\centering
\includegraphics[width=3in]{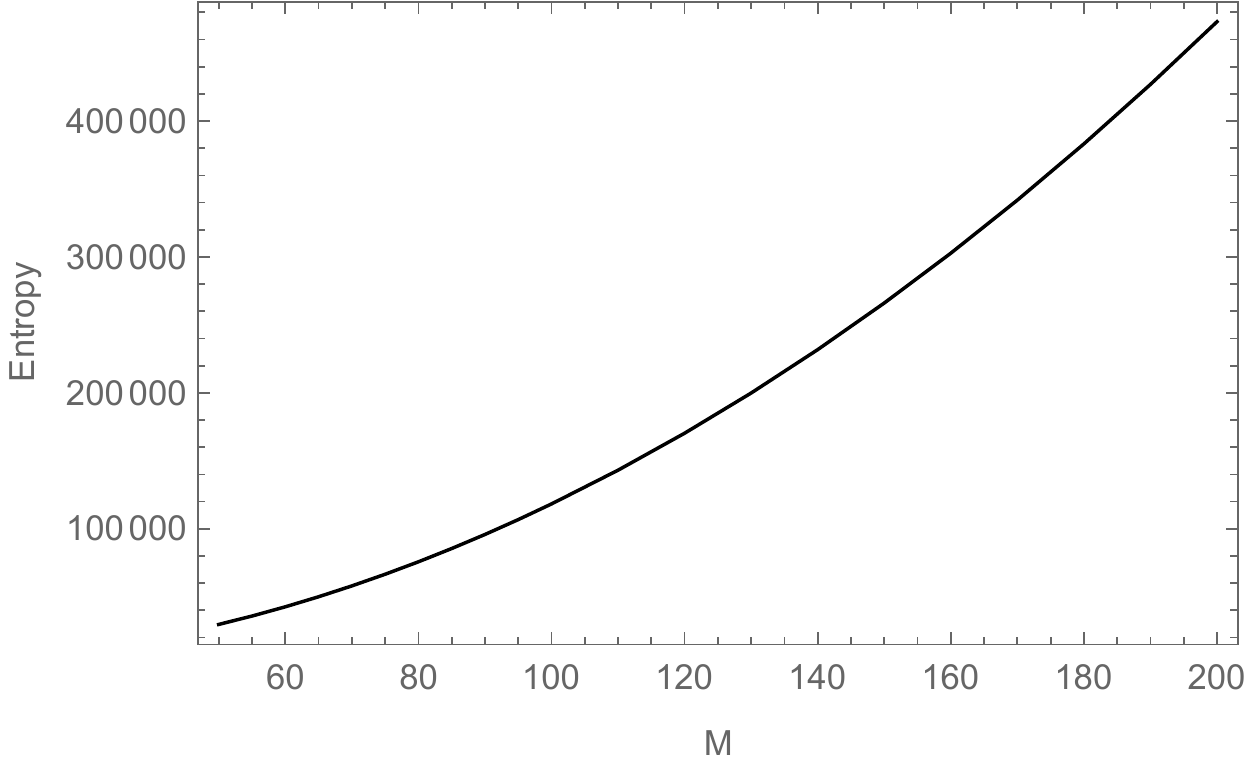}
\caption{Plot of the entropy against mass for the uncharged non-Kerr accelerating black hole for $a=0.5$, $\epsilon=0.5$, $\theta=\pi/2$, $\alpha=0.01.$}
\label{ent}
\end{figure}

Now a study of thermodynamic properties of black hole in the case of $q=0$ is made to compare with the results obtained for the charged case. For the spacetime (\ref{2.5}), the surface gravity and the Hawking temperature at the $r_{+}$ are
\begin{eqnarray}
\kappa_{+}&=&\frac{\Omega^{2}}{2(1+h)\rho^2}\Bigg\{(1-\alpha ^{2}r_{+}^{2})(2r_{+}-2M)-2\alpha^2 r_+ (a^2+r_{+}^2-2Mr_{+})\nonumber \\
&&+a^2\epsilon M^3 \sin^2\theta\left[\frac{1}{(r_{+}^{2}+a^2 \cos^2\theta)^2}-\frac{4r_{+}^{2}}{(r_{+}^{2}+a^2 \cos^2\theta)^3}\right]\Bigg\},\label{5.5}
\end{eqnarray}%
\begin{eqnarray}
T&=&\frac{\Omega^{2}}{4\pi(1+h)\rho^2}\Bigg\{(1-\alpha ^{2}r_{+}^{2})(2r_{+}-2M)-2\alpha^2 r_+ (a^2+r_{+}^2-2Mr_{+})\nonumber \\
&&+a^2\epsilon M^3 \sin^2\theta\left[\frac{1}{(r_{+}^{2}+a^2 \cos^2\theta)^2}-\frac{4r_{+}^{2}}{(r_{+}^{2}+a^2 \cos^2\theta)^3}\right]\Bigg\}.
\end{eqnarray}%
\begin{figure}[hptb]
\centering
\includegraphics[width=3in]{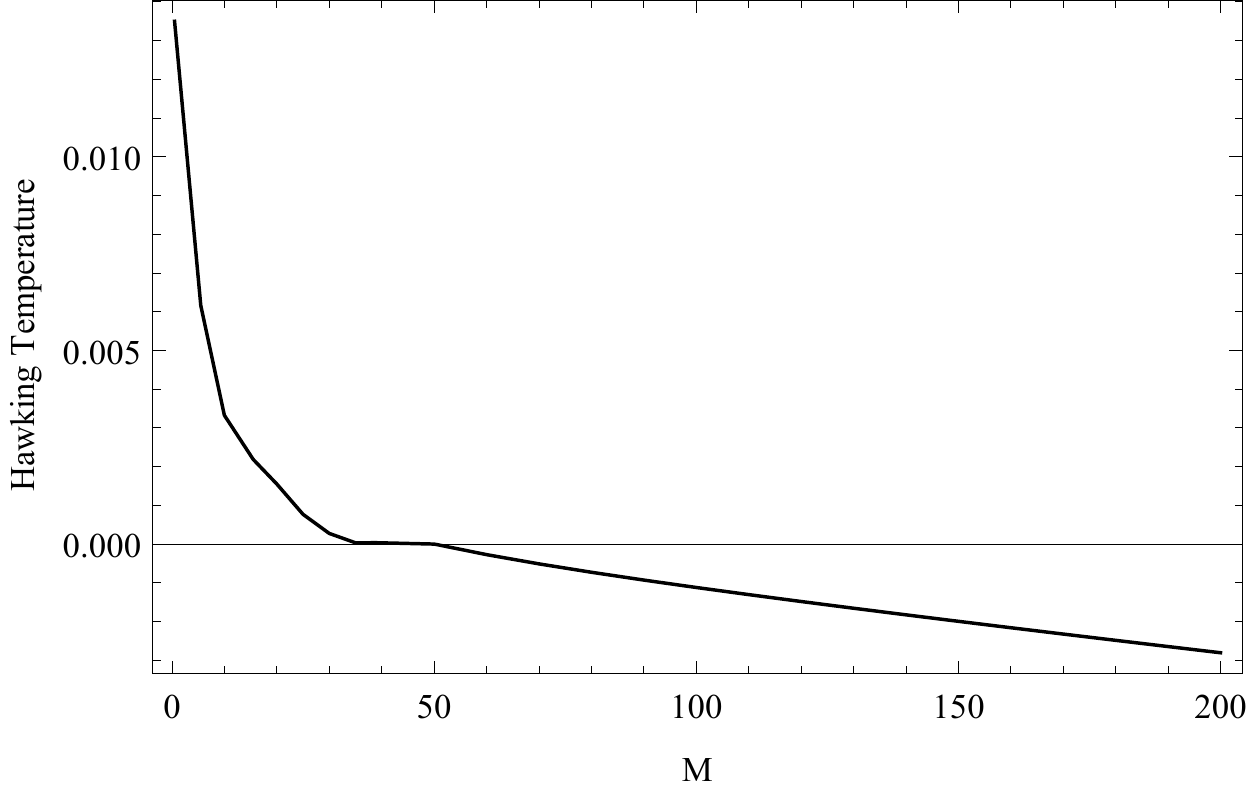}
\caption{Plot of the Hawking Temperature against mass for the uncharged non-Kerr accelerating black hole. Here, $a=0.5$, $\epsilon=0.5$, $\theta=\pi/2$, $\alpha=0.01.$}
\label{nt}
\end{figure}
The Hawking temperature at the equatorial plane is shown in Fig. \ref{nt}. Again we get positive and negative values. 
The angular velocity $\omega$, horizon area and the entropy for the uncharged non-Kerr accelerating black hole (\ref{2.5}) at the event horizon are
\begin{equation}
\omega_{+}=\frac{aP\left[(r_{+}^2+a^2)-(a^{2}-2Mr_{+}+r_{+}^{2})(1-\alpha ^{2}r_{+}^{2})\right]\left(1+h\right)}{P\left(r_{+}^2+a^2\right)^2-(a^{2}-2Mr_{+}+r_{+}^{2})(1-\alpha ^{2}r_{+}^{2})a^2 \sin^2\theta(1+h)}, \label{5.6}
\end{equation}
\begin{equation}
A=\frac{4\pi\left[P\left(r_{+}^2+a^2\right)^2-a^2 \sin^2\theta(a^{2}-2Mr_{+}+r_{+}^{2})(1-\alpha ^{2}r_{+}^{2})(1+h)\right] }{P\left[(r_{+}^2+a^2)-(a^{2}-2Mr_{+}+r_{+}^{2})(1-\alpha ^{2}r_{+}^{2})\right]\left(1+h\right)},
\end{equation}
\begin{equation}
S=\frac{\pi \left[P\left(r_{+}^2+a^2\right)^2-a^2 \sin^2\theta(a^{2}-2Mr_{+}+r_{+}^{2})(1-\alpha ^{2}r_{+}^{2})(1+h)\right]}{\left[P(r_{+}^2+a^2)-(a^{2}-2Mr_{+}+r_{+}^{2})(1-\alpha ^{2}r_{+}^{2})\right]\left(1+h\right)},
\end{equation}
where the value of $P$ is given in Eq. (\ref{2.5a}). Fig. \ref{ent} shows that entropy is an increasing function of $M$. The entropy behavior in Figs. \ref{cent} and \ref{ent} can be further elaborated from Table \ref{1}. It is clear from here that the presence of charge has led to a decrease in the entropy. 

\begin{table}[h!]
  \begin{center}
    \caption{Entropy for the non-Kerr accelerating black hole, for the charged and uncharged cases, for some values of mass $M$}
    \begin{tabular}{l|c|r} 
      $M$ & $S$ \text {(charged case)} & $S$ \text{(uncharged case)}\\
      \hline
      50 & 29623.8333 & 29624.6191\\
      55 & 35775.1506 & 35776.7594\\
      60 & 42575.5242 & 42577.1331\\
    65 & 49967.4241 & 49969.0330\\
70 & 57950.7245 & 57952.3333\\
    \end{tabular}
\label{1}
  \end{center}
\end{table}
In the first law of thermodynamics given in Eq. (\ref{4.4}), the last term vanishes as our spacetime is uncharged. By substituting the values of $\kappa_{+}$ and $\omega_{+}$ from Eqs. (\ref{5.5}) and (\ref{5.6}) in
(\ref{4.4}), the first law takes the form
\begin{align}
dm&=\frac{\Omega^{2}}{16\pi(1+h)\rho^2}\Bigg\{(1-\alpha ^{2}r_{+}^{2})(2r_{+}-2M)-2\alpha r_+ (a^2+r_{+}^2-2Mr_{+})+a^2\epsilon M \sin^2\theta\nonumber \\
&\left[\frac{1}{(r_{+}^{2}+a^2 \cos^2\theta)^2}-\frac{4r_{+}^{2}}{(r_{+}^{2}+a^2 \cos^2\theta)^3}\right]\Bigg\}dA+\frac{aP\left[(r_{+}^2+a^2)-Q\right]\left(1+h\right)}{P\left(r_{+}^2+a^2\right)^2{-}Qa^2 \sin^2\theta(1+h)} dJ.  \label{l2}
\end{align}
 The first law given in the Eq. (\ref{l1}) reduces to the Eq. (\ref{l2}) in case of vanishing charge. 
\section{Thermodynamics of the non-Kerr black hole}\label{thermojp}
\begin{figure}[!htpb]
\centering
\includegraphics[width=3.5in]{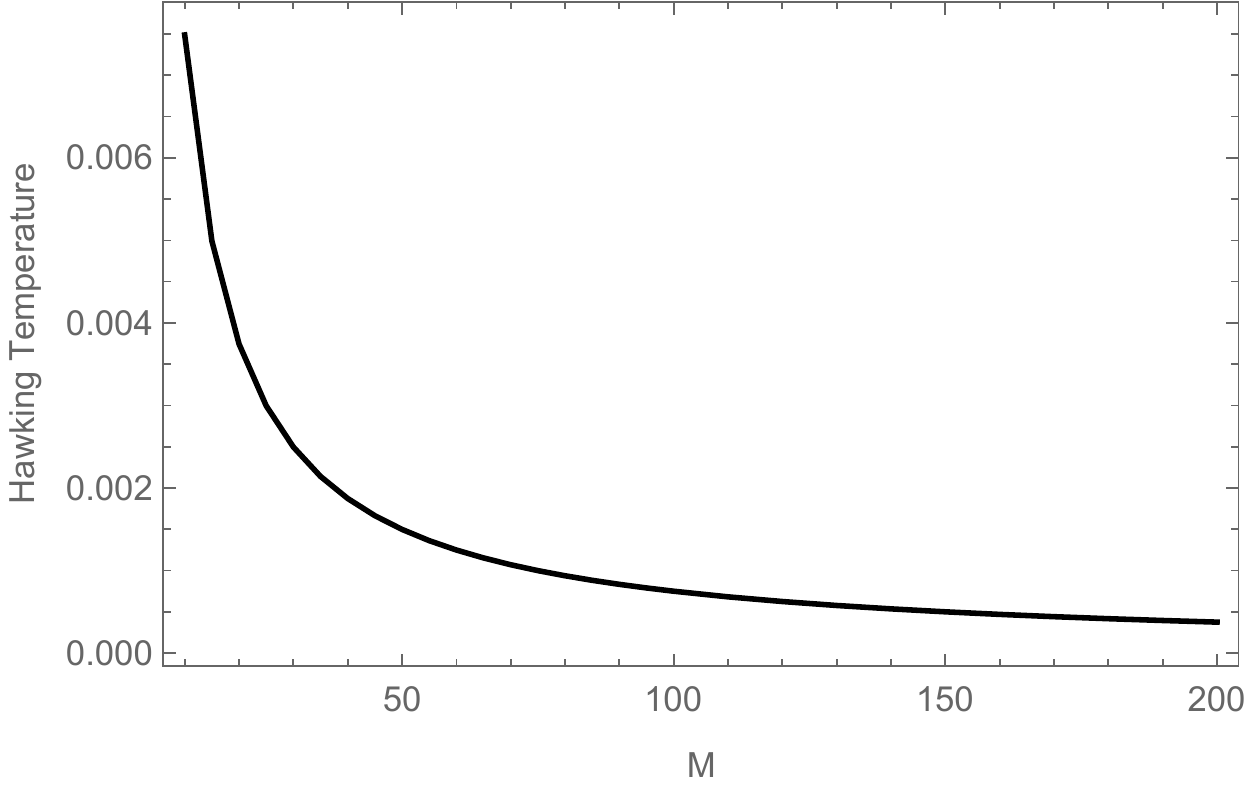}
\caption{Plot of the Hawking Temperature against mass for the non-Kerr black hole. Here we have taken $a=0.5$, $\epsilon=0.5$ and $\theta=\pi/2$.}\label{jpt}
\end{figure}

\begin{figure}[!hptb]
\centering
\includegraphics[width=3.5in]{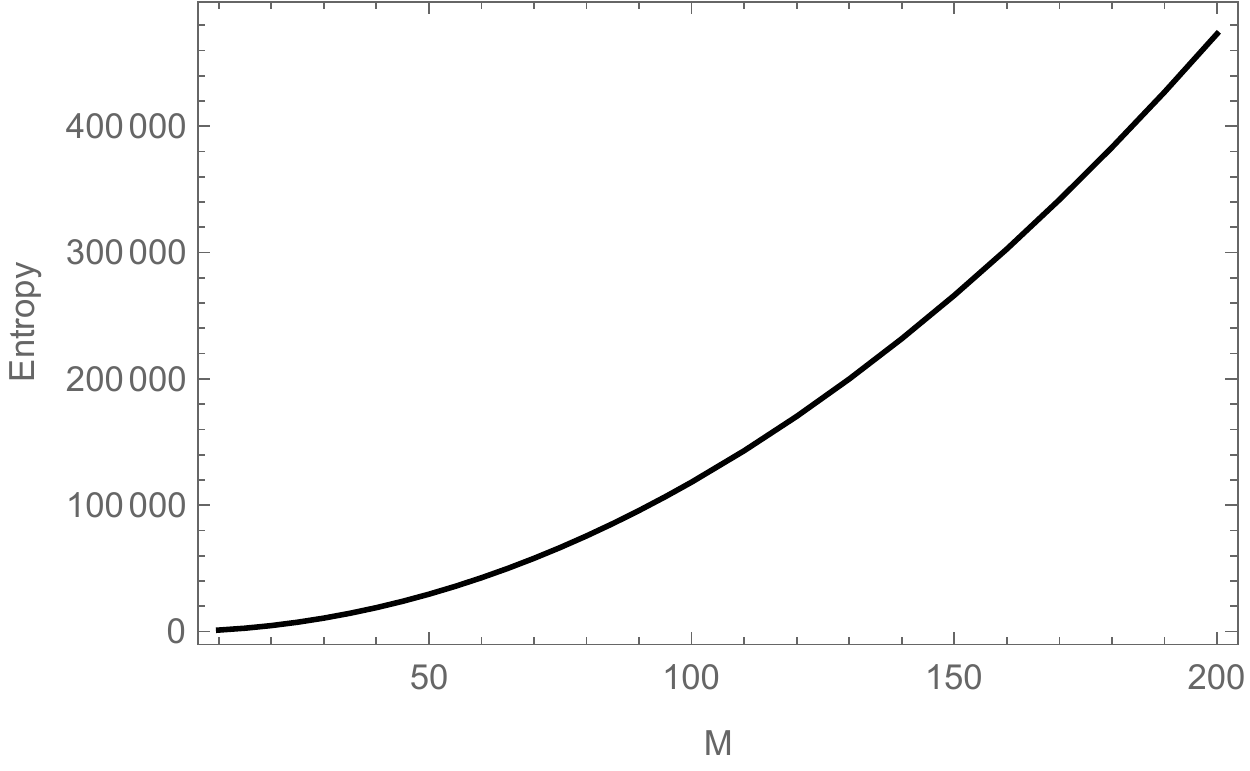}
\caption{Plot of the entropy against mass for the non-Kerr black hole. In this plot we have taken $a=0.5$, $\epsilon=0.5$ and $\theta=\pi/2$.}\label{sjp}
\end{figure}

For completeness and comparison, in this section, we work out the thermodynamic quantities for the non-Kerr spacetime  (\ref{2.3a}) with $q=0.$
The surface gravity and the Hawking temperature at $r_{+}$ are
\begin{eqnarray}
\kappa_{+}=\frac{1}{2(1+h)\rho^2}\Bigg\{(2r_{+}-2M)+a^2\epsilon M \sin^2\theta\left[\frac{1}{(r_{+}^{2}+a^2 \cos^2\theta)^2}-\frac{4r_{+}^{2}}{(r_{+}^{2}+a^2 \cos^2\theta)^3}\right]\Bigg\}.  \label{3.9a}
\end{eqnarray}%
\begin{eqnarray}
T=\frac{1}{2\pi(1+h)\rho^2}\Bigg\{(2r_{+}-2M)+a^2\epsilon M \sin^2\theta\left[\frac{1}{(r_{+}^{2}+a^2 \cos^2\theta)^2}-\frac{4r_{+}^{2}}{(r_{+}^{2}+a^2 \cos^2\theta)^3}\right]\Bigg\}.\nonumber \\
\end{eqnarray}%
The Hawking temperature has been plotted in Fig. \ref{jpt}. We see that when the acceleration parameter is absent, the negative values of temperature do not appear. 
The expressions for angular velocity, area and entropy at the event horizon are 
\begin{equation}
\omega_{+}=\frac{a\left[(r_{+}^2+a^2)-(a^{2}-2Mr_{+}+r_{+}^{2})\right]\left(1+h\right)}{\left(r_{+}^2+a^2\right)^2-a^2 \sin^2\theta(a^{2}-2Mr_{+}+r_{+}^{2})(1+h)}. \label{3.10a}
\end{equation}
\begin{equation}
A=\frac{4\pi \left(r_{+}^2+a^2\right)^2-a^2 \sin^2\theta(a^{2}-2Mr_{+}+r_{+}^{2})(1+h) }{\left[(r_{+}^2+a^2)-(a^{2}-2Mr_{+}+r_{+}^{2})\right]\left(1+h\right)}.
\end{equation}
\begin{equation}
S=\frac{\pi \left(r_{+}^2+a^2\right)^2-a^2 \sin^2\theta(a^{2}-2Mr_{+}+r_{+}^{2})(1+h) }{\left[(r_{+}^2+a^2)-(a^{2}-2Mr_{+}+r_{+}^{2})\right]\left(1+h\right)}.
\end{equation}
The entropy has been plotted in the Fig. \ref{sjp} which shows increasing behaviour with varying mass. 
The first law of thermodynamics in this case is
\begin{equation}
dm=\frac{\kappa_h}{8\pi} dA+ \omega_{+} dJ,
\end{equation}
\begin{eqnarray}
dm&=&\frac{1}{16\pi(1+h)\rho^2}\Bigg\{(2r_{+}-2M)+a^2\epsilon M \sin^2\theta\left[\frac{1}{(r_{+}^{2}+a^2 \cos^2\theta)^2}-\frac{4r_{+}^{2}}{(r_{+}^{2}+a^2 \cos^2\theta)^3}\right]\Bigg\} dA\nonumber \\
&&+ \frac{a\left[(r_{+}^2+a^2)-(a^{2}-2Mr_{+}+r_{+}^{2})\right]\left(1+h\right)}{\left(r_{+}^2+a^2\right)^2-a^2 \sin^2\theta(a^{2}-2Mr_{+}+r_{+}^{2})(1+h)} dJ.\nonumber \\
\end{eqnarray}.
We conclude this section on Table. \ref{1a} which highlights the plots of the Hawking temperature of this article in an effective way. It is observed that the presence of $\alpha$ not only decreases the temperature but is also responsible for the negative values of the temperature when compared with the case $\alpha=0=q$.
\begin{table}[h!]
  \begin{center}
    \caption{Hawking temperature of accelerating non-Kerr spacetimes}
    \begin{tabular}{l|c|c|r} 
      $M$ & $T_{H}$ \text {with } $\alpha\neq 0,\ q\neq 0$ & $T_{H}$ \text {with } $\alpha\neq 0,\ q=0$& $T_{H}$ \text {with } $\alpha=0,\ q= 0$\\
      \hline
      30 & 0.0006498 & 0.0002727 & 0.0012566\\
      35 & 0.0003953 & 0.0000304 & 0.0010770 \\
      40 & 0.0003367 & 0.0000324 & 0.0009362 \\
    45 & 0.0003016 & 0.0003369 & 0.0008203 \\
		50 & -0.0000036 & -0.0000264 & 0.0007489 \\
    \end{tabular}
\label{1a}
  \end{center}
\end{table}

\section{Conculsion}\label{dis}
In this work, we have formulated two deformed Kerr black hole spacetimes with acceleration and studied the effects of the deviation parameter $\epsilon$ and the acceleration parameter $\alpha$ on their structure and thermodynamics. The well-known accelerating black hole solution is the limiting case (when the deformation vanishes) of these new metrics.
  First, we have discussed their event horizon structure. We observe that the presence of the function $h(r,\theta)$ introduces special features into the event horizon which for accelerating black holes are represented by the same equation as in the Kerr-Newman and the Kerr solution for the charged and uncharged cases, respectively. The event horizons in our metrics showed spherical topology as well as the disjointed structure. The ergosphere for both the black holes is also discussed. 

In the analysis on thermodynamics, we observe that the presence of the acceleration parameter $\alpha$ makes the Hawking temperature negative for various mass values.
The negative value of the Hawking temperature has also been observed in literature. We have seen that the factor $(1-\alpha^2 r_{+}^2)$ makes the temperature negative for $\alpha>1/r_{+}$, restricting us to take only small values of acceleration \citep{kb, dl1, dl2}. The negative values of temperature thus obtained are extremely small. Therefore, to avoid the violation of the third law of black hole thermodynamics, it is required to limit $\alpha$ in the range $\alpha<1/r_{+}$. For small acceleration parameter, the temperature of our metrics showed positive behavior. 
The entropy profiles showed increasing values for both the spacetimes and thus satisfied the condition of the second law of thermodynamics.

The present work has also given an approach to study other kinds of deformations that are possible in the case of accelerating black holes. One may include other parameters like NUT charge and cosmological constant besides the acceleration parameter $\alpha$ to see the effect of deformations similar to the one presented here. 

\end{document}